\documentclass[a4paper,12pt]{article}
\usepackage{amsfonts}
\usepackage{latexsym}
\usepackage{amsmath}
\usepackage{amssymb}
\usepackage{amssymb}
\usepackage{slashed}
\usepackage{upgreek}
\usepackage{mathrsfs}
\usepackage{bbm}

\usepackage{amsthm}

\theoremstyle{remark}

\numberwithin{equation}{section}

\usepackage[utf8]{inputenc}
\usepackage[english]{babel}

\usepackage{amsmath}
\usepackage{amssymb} 
\usepackage{slashed}
\usepackage{mathtools} 
\usepackage{dsfont} 
\usepackage{xcolor}

\usepackage[toc,page]{appendix}

\renewcommand{\*}[1]{{\,^* \!#1}} 
\newcommand{\dd}{\mathrm{d}}

\newcommand{\pp}{=\kern-0.40em{\vert}}

\theoremstyle{definition}

\newcommand*{\defeq}{\mathrel{\vcenter{\baselineskip0.5ex \lineskiplimit0pt
                     \hbox{\scriptsize.}\hbox{\scriptsize.}}}%
                     =}

\allowdisplaybreaks

\usepackage{relsize}
\usepackage{graphicx}
\usepackage{comment}

\renewcommand{\thefootnote}{\fnsymbol{footnote}}

\def\appendix#1{\addtocounter{section}{1}\setcounter{equation}{0}
\renewcommand{\thesection}{\Alph{section}}
\section*{Appendix \thesection\protect\indent \parbox[t]{11.15cm}{#1}}
\addcontentsline{toc}{section}{Appendix \thesection\ \ \ #1}}

\def\bbe{{\bf{e}}}

\font\mybb=msbm10 at 11pt

\def\bb#1{\hbox{\mybb#1}}

\def\bR {\bb{R}}

\def\bC {\bb{C}}

\newcommand{\bea}{\begin{eqnarray}}
\newcommand{\eea}{\end{eqnarray}}
\newcommand{\nn}{\nonumber \\}

\def\rep{\operatorname{Re}}
\def\imp{\operatorname{Im}}

\def\HH{\bar{H}}
\def\PP{\bar{\Phi}}
\def\GG{\bar{G}}
\def\dd{\mathrm{d}}
\def\ee{\boldsymbol{\mathrm{e}}}
\usepackage{mathtools}

\usepackage{centernot}
\newcommand{\fsl}[1]{{\centernot{#1}}}
\def\rrho{\mathring{\rho}}
\def\romega{\mathring{\omega}}
\def\trrho{\mathring{\tilde{\rho}}}
\def\tromega{\mathring{\tilde{\omega}}}
\def\kkappa{\mathring{\kappa}}
\def\tkkappa{\mathring{\tilde{\kappa}}}

\usepackage[normalem]{ulem}				
\usepackage{microtype}
\usepackage[colorlinks, citecolor=blue, linkcolor=blue, urlcolor=Maroon, filecolor=Maroon, linktocpage=true]{hyperref} 

\usepackage{chngcntr}

\begin{document}

\begin{center}
\vspace*{-1.0cm}
\begin{flushright}
\end{flushright}


\vspace{2.0cm} {\Large \bf TCFHs, IIB warped AdS  backgrounds and hidden symmetries } \\[.2cm]

\vskip 2cm
 L.~ Grimanellis and G.~ Papadopoulos
\\
\vskip .6cm

\begin{small}
\textit{Department of Mathematics
\\
King's College London
\\
Strand
\\
 London WC2R 2LS, UK}\\
\texttt{loukas.grimanellis@kcl.ac.uk}
\\
\texttt{george.papadopoulos@kcl.ac.uk}
\end{small}
\\*[.6cm]

\end{center}

\vskip 2.5 cm

\begin{abstract}
\noindent

We present the twisted covariant form hierarchies (TCFHs) on the internal spaces  of all  type IIB warped AdS backgrounds. As a result we demonstrate that the form bilinears on the internal spaces satisfy a generalisation of the conformal Killing-Yano equation.  We also explore some of the properties of the TCFHs, like for example the holonomy of the TCFH connections.  In addition, we present examples where the form bilinears generate hidden symmetries for particle probes propagating on the internal spaces of some AdS backgrounds.  These  include the maximally supersymmetric AdS$_5$ solution as well as  some of the  near horizon geometries of  intersecting IIB branes.

\end{abstract}

\vskip 1.5 cm


\newpage

\renewcommand{\thefootnote}{\arabic{footnote}}


\section{Introduction}

Recently it has been demonstrated that the conditions imposed on the Killing spinor form bilinears, as a consequence of the gravitino Killing spinor equation (KSE) of any supergravity theory\footnote{The supergravity theory may include higher curvature corrections and be defined on a spacetime of any signature.}, can be organised as a twisted covariant form hierarchy (TCFH) \cite{jggp, tcfhgp}.  This means that there is a connection ${\cal D}^{{\cal F}}$ on a suitable space of spacetimes forms such that schematically
\bea
{\cal D}^{\cal F}_X \Omega=i_X {\cal P}+ X \wedge {\cal Q}~,
\label{tcfhg}
\eea
for any spacetime vector field $X$, where $\Omega$ is a multi-form spanned by the form bilinears, and ${\cal P}$ and ${\cal Q}$ are multi-forms that depend on the form bilinears and the (form) fluxes ${\cal F}$ of the theory. The TCFH connection ${\cal D}^{\cal F}$ is not necessarily form degree preserving. A consequence of the existence of the TCFHs is that the form bilinears of all supergravity theories satisfy a generalisation of the conformal Killing-Yano\footnote{The standard CKY condition on a k-form $\omega$ is $\nabla_X\omega=i_X d\omega-{1\over n-k+1} X\wedge \delta\omega$, where $\nabla$ is the Levi-Civita connection of a metric $g$.  If $\omega$ is co-closed, $\delta\omega=0$, then $\omega$ is a Killing-Yano (KY) form, while, if $\omega$ is closed, then $\omega$ is a closed CKY (CCKY) form.}  (CKY) equation with respect to ${\cal D}^{\cal F}$.

It is well-known that KY forms are associated with conservation laws of the geodesic flow and the integrability of some classical field equations on some black hole spacetimes \cite{penrose}-\cite{lun}, see also reviews \cite{revky, frolov} and references therein. They also generate symmetries \cite{gibbons} for  spinning particles probes \cite{bvh} propagating on a spacetime.  For other applications, see \cite{gggpks}-\cite{ls2}. Therefore, it is natural  to raise the question on whether the form bilinears generate symmetries for various particle probes propagating on supersymmetric spacetimes. Much partial progress has been made to answer this question in \cite{ebgp1, ebgp2, lgjpgp, ebgp3}.

In this paper, we shall demonstrate that the conditions imposed on the Killing spinor form bilinears on the internal space of all IIB AdS backgrounds by the gravitino KSE of the theory can be organised as a TCFH.  In particular, we shall determine the TCFH connection ${\cal D}^{\cal F}$ and investigate some of its properties like its (reduced) holonomy on generic backgrounds. In addition, we demonstrate that the form bilinears of some AdS backgrounds, which include the maximally supersymmetric AdS$_5$ solution as well as the near horizon geometries of some intersecting brane configurations, are either KY or CCKY\footnote{The Hodge dual of a CCKY form is a KY form and vice versa.} forms and therefore generate symmetries for some spinning particle probes propagating on the internal space of these backgrounds.

This paper is organised as follows. In sections 2, 3, 4 and 5, we present the TCFHs on the internal spaces of AdS$_k$ backgrounds, $k\geq 2$, and describe some of the properties of their TCFH connections. In section 6, we present some  examples of AdS backgrounds whose Killing spinor form bilinears generate symmetries for spinning particle probes, and in section 7 we give our conclusions.  In appendix A, we describe our conventions. In appendix B, we prove the Liouville integrability of geodesic flow on all AdS$_k\times S^m \times \bR^n$ backgrounds, and in appendix C we give the TCFH of IIB supergravity in the Einstein frame.

\section{The TCFH of warped  AdS$_2$ backgrounds}

\subsection{Fields and Killing spinor equations}\label{s1sub1}

Let $g$  be the spacetime metric,  and $G$, $F$ and $P$ be the $U(1)$-twisted 3-form, 5-form and $U(1)$-twisted 1-form field strengths of IIB supergravity \cite{js} in the Einstein frame, respectively.  These fields for warped AdS$_2$ backgrounds, AdS$_2\times_w N^8$,  can be expressed \cite{sbjggp} as
\begin{gather}
g = 2\,\ee^+\ee^-  + g(N^8)~, \nn
F = \ee^+ \wedge \ee^- \wedge Y + {}^{\star_8}Y~, \quad G = \ee^+ \wedge \ee^- \wedge \Phi + H~, \quad P=\xi~,
\end{gather}
where $g(N^8)$ is a metric on $N^8$,  $Y$ is a 2-form on the internal space $N^8$,  $\Phi$ and $\xi$ are $U(1)$-twisted 1-forms and   $H$ is a $U(1)$-twisted 3-form on $N^8$. The pseudo-orthonormal frame, $(\bbe^+, \bbe^-, \bbe^i)$,  on the spacetime is expressed as
\bea
\ee^+ =  du~, \quad \ee^- = dr+rh-\frac{1}{2}r^2\ell^{-2}A^{-2} \,du~, \quad \ee^i = e^i_I\, dy^I~,
\eea
with $\bbe^i$ an orthonormal frame on $N^8$, $g(N^8)=\delta_{ij} \bbe^i \bbe^j$, and $h=-2 A^{-1} d A$, where $A$ is the warped factor,  $y^I$ are the coordinates of $N^8$ and $(u,r)$ are the remaining spacetime coordinates. It can be seen after a coordinate transformation that the spacetime metric can be
written in the standard warped form $g=A^2 g_\ell (AdS_2)+ g(N^8)$, where $g_\ell (AdS_2)$ is the standard metric on AdS$_2$ with radius $\ell$.

 The gravitino and dilatino Killing spinor equations (KSEs) of IIB supergravity can be integrated over the coordinates $(u,r)$ \cite{sbjggp}. One finds that the Killing spinors $\epsilon$ can be expressed as $\epsilon=\epsilon(u, r, \eta_\pm)$, where\footnote{From here on, all the gamma matrices are taken with respect to a spacetime pseudo-orthonormal frame as that stated above.}   $\Gamma_\pm \eta_\pm=0$ and $\eta_\pm$ depend only on the coordinates $y$ of $N^8$.
 In addition, as a consequence of the gravitino KSE of the theory,  one finds that $\eta_\pm$  satisfy the KSEs
\bea \label{KSE_eta}
\nabla^{(\pm)}_i\eta_{\pm}=0~,
\eea
on $N^8$, where
 the supercovariant derivatives are
\begin{align}\label{KSE2}
\nabla_{i}^{(\pm)} \equiv \nabla_i &+ \left( -\frac{i}{2}\,Q_i \pm \frac{1}{2}\,\partial_i \log A \mp \frac{i}{4}\, \slashed{Y}_i \pm \frac{i}{12}\,(\Gamma\slashed{Y})_i \right) \nn
&+\left( \pm \frac{1}{16}\,(\Gamma\slashed{\Phi})_i \mp\frac{3}{16}\,\Phi_i -\frac{1}{96}\,(\Gamma\slashed{H})_i +\frac{3}{32}\,\slashed{H}_i \right)C*~,
\end{align}
  $\nabla$ is the connection induced on the spin bundle from the Levi-Civita connection of $g(N^8)$, and the anti-linear operation\footnote{We follow the spinor conventions of \cite{Gran:2018ijr} appendix B, see also appendix A. In the basis of that paper  $C=\Gamma_{6789}$.}
   $C*$ commutes with all the gamma matrices and squares to the identity map, i.e. $C*$ can be used as a spin invariant reality condition. $Q$ is a $U(1)$ connection on $N^8$ constructed from the scalar fields of IIB theory. The spinor $\eta_\pm$ satisfy additional condition on $N^8$ arising from the dilatino KSE of IIB supergravity. These condition will be explored later in examples that we shall present but they are not essential in the investigation of the TCFH of the warped AdS$_2$ backgrounds.

\subsection{The TCFH and holonomy }

To present the TCFH of AdS$_2$ backgrounds consider some spinors  $\eta^r_{\pm}$, $r=1, \dots, N/2$,  and construct a basis\footnote{The bases in the space of form bilinears that we are considering are up to a Hodge duality operation on the internal space.} in the space of form bilinears on the internal space $N^8$ as
\begin{gather}
\rho^{rs}_{\pm} = \langle \eta^r_{\pm}, \eta^s_{\pm} \rangle~, \quad \tilde{\rho}^{rs}_{\pm} = \langle \eta^r_{\pm}, C\bar{\eta}^s_{\pm} \rangle~, \nn
\omega^{rs}_{\pm} = \frac{1}{2}\, \langle \eta^r_{\pm}, \Gamma_{i_1 i_2}\, \eta^s_{\pm} \rangle\, \ee^{i_1}\wedge \ee^{i_2}~, \quad \tilde{\omega}^{rs}_{\pm} = \frac{1}{2}\, \langle \eta^r_{\pm}, \Gamma_{i_1 i_2}\, C\bar{\eta}^s_{\pm} \rangle\, \ee^{i_1}\wedge \ee^{i_2} ~, \nn
\zeta^{rs}_{\pm} = \frac{1}{4!}\, \langle \eta^r_{\pm}, \Gamma_{i_1\dots i_4}\, \eta^s_{\pm} \rangle\, \ee^{i_1}\wedge\dots \wedge \ee^{i_4}~, \quad \tilde{\zeta}^{rs}_{\pm} = \frac{1}{4!}\, \langle \eta^r_{\pm}, \Gamma_{i_1\dots i_4}\, C\bar{\eta}^s_{\pm} \rangle \, \ee^{i_1}\wedge\dots \wedge \ee^{i_4} ~.
\label{ads2_bilinears}
\end{gather}
where $C*\eta_{\pm}=C\bar{\eta}_{\pm}$, with $\bar{\eta}_{\pm}$ the complex conjugate of $\eta_{\pm}$ and (time-) space-like gamma matrices are (anti-)Hermitian with respect to the inner product $\left\langle\cdot, \cdot \right\rangle$. In fact $\tilde\rho$, $\tilde\omega$ and $\tilde \zeta$ are $U(1)$-twisted forms on $N^8$.  Moreover, $\zeta_+$ is self-dual, while $\zeta_-$ is anti-self-dual, on $N^8$, and similarly for $\tilde \zeta_+$ and $\tilde \zeta_-$. This is a consequence of the chirality of $\eta_\pm$ as IIB spinors and the
conditions $\Gamma_\pm\eta_\pm=0$  which in turn imply that  $(\prod_{i=1}^8\Gamma_i) \eta_\pm=\pm\eta_\pm$. Furthermore, $\rep\rho_\pm^{rs}$, $\imp\omega_\pm^{rs}$, $\rep\zeta_\pm^{rs}$,  $\tilde{\rho}^{rs}_{\pm}$ and $\tilde{\zeta}^{rs}_{\pm}$ are symmetric, while $\imp\rho_\pm^{rs}$, $\rep\omega_\pm^{rs}$, $\imp\zeta_\pm^{rs}$ and $\tilde{\omega}^{rs}_{\pm}$ are skew-symmetric, in the exchange of the spinors $\eta^r_{\pm}$ and $\eta^s_{\pm}$.

Assuming that $\eta^r_{\pm}$ are Killings spinors on $N^8$, i.e. allowing $\eta_\pm^r$ to satisfy (\ref{KSE_eta}), and using the identity
\bea
\nabla_i\,\phi^{rs}_{\pm i_1\dots i_k} = \left\langle \nabla_i\,\eta_{\pm}^r, \Gamma_{i_1\dots i_k}\eta_{\pm}^s \right\rangle + \left\langle \eta_{\pm}^r, \Gamma_{i_1\dots i_k}\nabla_i\,\eta_{\pm}^s \right\rangle~,
\eea
where $\phi$ stands for any of the form blinears above, one finds after some extensive Clifford algebra computation that
 \bea
{\cal D}^{\cal (\pm)F}_i \rho^{rs}_{\pm} &\defeq& \nabla_i \rho^{rs}_{\pm} = \mp \partial_i \log A \rho^{rs}_{\pm} \pm \frac{i}{2}\, Y_i{}^{j_1j_2}\,\omega^{rs}_{\pm j_1j_2} \pm \frac{3}{8}\,\rep \{\Phi_i \tilde{\rho}^{rs}_{\pm}\}  \cr
&&+\frac{1}{48}\, \rep \{ H^{j_1j_2j_3} \,\tilde{\zeta}^{rs}_{\pm i j_1j_2j_3}\} \mp\frac{i}{8}\,\imp\{\Phi^j\,\tilde{\omega}^{rs}_{\pm ij}\} \cr
&&- \frac{3i}{16}\,\imp\{H_i{}^{j_1j_2}\,\tilde{\omega}^{rs}_{\pm j_1j_2}\} ~,
\label{ads2tcfh1}
\eea
\bea
{\cal D}^{\cal (\pm)F}_i \omega^{rs}_{\pm i_1 i_2} &\defeq& \nabla_i\, \omega^{rs}_{i_1 i_2} \pm \partial_i \log A \, \omega^{rs}_{\pm i_1 i_2} \mp i\,Y^{j_1j_2}{}_i\,\zeta^{rs}_{\pm i_1i_2j_1j_2} +\frac{i}{4}\, \imp\{H_i{}^{j_1j_2} \tilde{\zeta}^{rs}_{\pm i_1i_2j_1j_2}\} \cr
&& \mp\frac{1}{2}\,\rep\{\Phi_i\,\tilde{\omega}^{rs}_{\pm i_1i_2}\} - \rep\{H^j{}_{i[i_1}\,\tilde{\omega}^{rs}_{\pm i_2]j}\} \cr
&&= \mp i\,Y_{ii_1i_2} \rho^{rs}_{\pm} \pm \frac{i}{3}\,Y^{j_1j_2j_3}\,\delta_{i[i_1}\,\zeta^{rs}_{\pm i_2]j_1j_2j_3} \mp \frac{3i}{2} \, Y^{j_1j_2}{}_{[i}\,\zeta^{rs}_{\pm i_1i_2]j_1j_2} \cr
&&\mp \frac{i}{8}\,\imp \{\Phi^j \,\tilde{\zeta}^{rs}_{\pm i i_1 i_2j}\} \pm \frac{i}{4}\, \delta_{i[i_1}\,\imp\{\Phi_{i_2]}\tilde{\rho}^{rs}_{\pm}\} -\frac{i}{24}\,\delta_{i[i_1}\imp\{H^{j_1j_2j_3}\,\tilde{\zeta}^{rs}_{\pm i_2]j_1j_2j_3}\}  \cr
&&+ \frac{3i}{16}\, \imp\{H^{j_1j_2}{}_{[i_1}\tilde{\zeta}^{rs}_{\pm i_2i]j_1j_2}\} +\frac{3i}{8}\, \imp\{H_{ii_1i_2}\tilde{\rho}^{rs}_{\pm}\} \pm\frac{1}{4}\,\delta_{i[i_1}\rep\{\Phi^j\,\tilde{\omega}^{rs}_{\pm i_2]j}\} \cr
&&\mp\frac{3}{8}\,\rep\{\Phi_{[i_1}\,\tilde{\omega}^{rs}_{\pm i_2i]}\} + \frac{1}{16}\,\rep\{{}^{\star}H_{ii_1i_2}{}^{j_1j_2}\,\tilde{\omega}^{rs}_{\pm j_1j_2}\} - \frac{1}{8}\,\delta_{i[i_1}\,\rep\{H_{i_2]}{}^{j_1j_2}\,\tilde{\omega}^{rs}_{\pm j_1j_2}\} \cr
&&-\frac{3}{8}\,\rep\{H^j{}_{[i_1i_2}\,\tilde{\omega}^{rs}_{\pm i]j}\} ~, \nn
\label{ads2tcfh2}
\eea
\bea
{\cal D}^{\cal (\pm)F}_i \zeta^{rs}_{\pm i_1\dots i_4} &\defeq& \nabla_i\,\zeta^{rs}_{\pm i_1\dots i_4} \pm \partial_i \log A\, \zeta^{rs}_{\pm i_1\dots i_4} - 8i\, {}^{\star}Y^j{}_{i[i_1i_2i_3}\omega^{rs}_{\pm i_4]j} \pm 12i\, Y_{i[i_1i_2}\,\omega^{rs}_{\pm i_3i_4]} \cr
&&\mp \frac{1}{2}\, \rep\{ \Phi_i \,\tilde{\zeta}^{rs}_{\pm i_1\dots i_4}\} -2\rep\{H^j{}_{i[i_1}\tilde{\zeta}^{rs}_{\pm i_2i_3i_4]j}\} +2i\,\imp\{{}^{\star}H^j{}_{i[i_1i_2i_3}\,\tilde{\omega}^{rs}_{\pm i_4]j}\} \cr
&& -3i\,\imp\{H_{i[i_1i_2}\,\tilde{\omega}^{rs}_{\pm i_3i_4]}\} \cr
&&= -2i\, \delta_{i[i_1}\,{}^{\star}Y_{i_2i_3i_4]}{}^{j_1j_2}\omega^{rs}_{\pm j_1j_2} -5i \,{}^{\star}Y^j{}_{[i_1\dots i_4}\omega^{rs}_{\pm i]j} \mp 12i\,\delta_{i[i_1}\,Y_{i_2i_3}{}^j\,\omega^{rs}_{\pm i_4]j} \cr
&&\pm 10i\, Y_{[i_1i_2i_3}\,\omega^{rs}_{\pm i_4i]} \pm \frac{1}{2}\,\delta_{i[i_1}\rep\{\Phi^j\,\tilde{\zeta}^{rs}_{\pm i_2i_3i_4]j}\} \mp \frac{5}{8}\rep\{\Phi_{[i}\,\tilde{\zeta}^{rs}_{\pm i_1\dots i_4]}\} \cr
&&\mp \frac{1}{8}\rep\{{}^{\star}H_{ii_1\dots i_4} \tilde{\rho}^{rs}_{\pm}\} -\frac{3}{4}\,\delta_{i[i_1} \rep\{H_{i_2}{}^{j_1j_2}\,\tilde{\zeta}^{rs}_{\pm i_3 i_4] j_1j_2}\} -\frac{5}{4}\,\rep\{H^j{}_{[i_1i_2}\,\tilde{\zeta}^{rs}_{\pm i_3i_4i]j}\} \cr
&&+\frac{1}{2}\, \delta_{i[i_1}\,\rep\{ H_{i_2i_3i_4]}\tilde{\rho}^{rs}_{\pm}\} \mp \frac{i}{16}\,\imp\{{}^{\star}\Phi_{i_1\dots i_4i}{}^{j_1j_2}\,\tilde{\omega}^{rs}_{\pm j_1j_2}\} \pm\frac{3i}{2}\,\delta_{i[i_1}\,\imp\{\Phi_{i_2}\,\tilde{\omega}^{rs}_{\pm i_3i_4]}\} \cr
&&+\frac{3i}{2}\,\delta_{i[i_1}\,\imp\{H_{i_2i_3}{}^j\,\tilde{\omega}^{rs}_{\pm i_4]j}\} -\frac{5i}{4}\,\imp\{H_{[i_1i_2i_3}\,\tilde{\omega}^{rs}_{\pm i_4i]}\} \cr
&&+ \frac{3i}{4}\,\delta_{i[i_1}\,\imp\{{}^{\star}H_{i_2i_3i_4]}{}^{j_1j_2}\,\tilde{\omega}^{rs}_{\pm j_1j_2}\} +\frac{15i}{8}\,\imp\{{}^{\star}H^j{}_{[i_1i_2i_3i_4}\,\tilde{\omega}^{rs}_{\pm i]j}\} ~,
\label{ads2tcfh3}
\eea
\bea
{\cal D}^{\cal (\pm)F}_i \tilde{\rho}^{rs}_{\pm} &\defeq& \nabla_i\tilde{\rho}^{rs}_{\pm} +(i Q_i \pm \partial_i \log A)\,\tilde{\rho}^{rs}_{\pm}= \pm \frac{i}{6}\, Y^{j_1j_2j_3}\,\tilde{\zeta}^{rs}_{\pm i j_1j_2j_3} \mp \frac{1}{8}\,\PP^j\,\omega^{(rs)}_{\pm ij} \pm \frac{3}{8}\, \PP_i \rho^{(rs)}_{\pm} \cr
&& + \frac{1}{48} \, \HH^{j_1j_2j_3}\zeta^{(rs)}_{\pm ij_1j_2j_3} - \frac{3}{16}\, \HH_i{}^{j_1j_2}\omega^{(rs)}_{\pm j_1j_2}~,
\label{ads2tcfh4}
\eea
\bea
{\cal D}^{\cal (\pm)F}_i\tilde{\omega}^{rs}_{\pm i_1i_2} &\defeq& \nabla_i\,\tilde{\omega}^{rs}_{\pm i_1i_2} +(i Q_i \pm \partial_i\log A)\,\tilde{\omega}^{rs}_{\pm i_1i_2} \mp 4i\, Y^j{}_{i[i_1}\,\tilde{\omega}^{rs}_{\pm i_2]j} \mp\frac{1}{2}\,\PP_i \omega^{[rs]}_{\pm i_1i_2} \cr
&&+\frac{1}{4}\,\HH_i{}^{j_1j_2}\,\zeta^{[rs]}_{\pm i_1i_2j_1j_2} -\HH^{j}{}_{i[i_1}\,\omega^{[rs]}_{\pm i_2]j} \cr
&&= \pm \frac{i}{2}\,{}^{\star}Y_{ii_1i_2}{}^{j_1j_2}\,\tilde{\omega}^{rs}_{\pm j_1j_2} \mp i\,\delta_{i[i_1}\,Y_{i_2]}{}^{j_1j_2}\,\tilde{\omega}^{rs}_{\pm j_1j_2} \mp 3i\,Y^j{}_{[i_1i_2}\,\tilde{\omega}^{rs}_{\pm i]j} \cr
&&\mp\frac{1}{8}\,\PP^j\zeta^{[rs]}_{\pm i_1i_2ij} \pm \frac{1}{4}\,\PP^j\,\delta_{i[i_1}\omega^{[rs]}_{\pm i_2]j} \mp\frac{3}{8}\,\PP_{[i_1}\,\omega^{[rs]}_{\pm i_2i]} \pm\frac{1}{4}\,\delta_{i[i_1}\,\PP_{i_2]}\,\rho^{[rs]}_{\pm} \cr
&&+\frac{1}{16}\,{}^{\star}\HH_{ii_1i_2}{}^{j_1j_2}\,\omega^{[rs]}_{\pm j_1j_2} -\frac{1}{24}\,\HH^{j_1j_2j_3}\,\delta_{i[i_1}\,\zeta^{[rs]}_{\pm i_2] j_1j_2j_3} +\frac{3}{16}\,\HH^{j_1j_2}{}_{[i_1}\,\zeta^{[rs]}_{\pm i_2i] j_1j_2} \cr
&&-\frac{1}{8}\,\delta_{i[i_1}\,\HH_{i_2]}{}^{j_1j_2}\, \omega^{[rs]}_{\pm j_1j_2} -\frac{3}{8}\,\HH^j_{[i_1i_2}\,\omega^{[rs]}_{\pm i]j} + \frac{3}{8}\,\HH_{ii_1i_2}\,\rho^{[rs]}_{\pm} ~,
\label{ads2tcfh5}
\eea
\bea
{\cal D}^{\cal (\pm)F}_i \tilde{\zeta}^{rs}_{\pm i_1\dots i_4} &\defeq& \nabla_i \,\tilde{\zeta}^{rs}_{\pm i_1\dots i_4} + (i Q_i \pm \partial_i \log A)\,\tilde{\zeta}^{rs}_{\pm i_1\dots i_4} \mp 8i\, Y^j{}_{i[i_1}\,\tilde{\zeta}^{rs}_{\pm i_2i_3i_4]j} \mp \frac{1}{2}\, \PP_i\, \zeta^{(rs)}_{\pm i_1 \dots i_4} \nn
&&\pm 2{}^{\star}\HH^j{}_{i[i_1i_2i_3}\omega^{(rs)}_{\pm i_4]j}-2\HH^j_{i[i_1}\,\zeta^{(rs)}_{\pm i_2i_3i_4]j} -3\, \HH_{i[i_1i_2} \omega^{(rs)}_{\pm i_3 i_4]} \cr
&&=-i\,{}^{\star} Y_{ii_1\dots i_4}\tilde{\rho}^{rs}_{\pm} \mp 6i\,\delta_{i[i_1} Y_{i_2}{}^{j_1j_2}\,\tilde{\zeta}^{rs}_{\pm i_3i_4]j_1j_2} \mp 10i\, Y^j_{}{[i_1i_2}\,\tilde{\zeta}^{rs}_{\pm i_3i_4i]j} \cr
&&\pm 4i\, \delta_{i[i_1}Y_{i_2i_3i_4]}\,\tilde{\rho}^{rs}_{\pm} - \frac{1}{16}{}^{\star}\PP_{ii_1\dots i_4}{}^{j_1j_2}\,\omega^{(rs)}_{\pm j_1j_2} \pm \frac{1}{2}\PP^j\, \delta_{i[i_1}\, \zeta^{(rs)}_{\pm i_2 i_3 i_4]j} \cr
&&\mp \frac{5}{8} \,\PP_{[i_1}\,\zeta^{(rs)}_{\pm i_2i_3i_4i]} \pm \frac{3}{2}\, \delta_{i[i_1}\PP_{i_2}\,\omega^{(rs)}_{\pm i_3i_4]} \mp\frac{1}{8}{}^{\star}\HH_{ii_1\dots i_4}\, \rho^{(rs)}_{\pm} \cr
&&-\frac{3}{4}\,\delta_{i[i_1}\HH_{i_2}{}^{j_1j_2}\,\zeta^{(rs)}_{\pm i_3i_4]j_1j_2} - \frac{5}{4}\,\HH^j{}_{[i_1i_2}\,\zeta^{(rs)}_{\pm i_3i_4i]j} +\frac{3}{2}\,\delta_{i[i_1}\HH_{i_2i_3}{}^j\omega^{(rs)}_{\pm i_4]j} \cr
&&- \frac{5}{4}\,\HH_{[i_1i_2i_3}\,\omega^{(rs)}_{\pm i_4 i]} +\frac{1}{2}\,\delta_{i[i_1}\HH_{i_2i_3i_4]}\, \rho^{(rs)}_{\pm} \pm \frac{3}{4}\, \delta_{i[i_1}{}^{\star} \HH_{i_2i_3i_4]}{}^{j_1j_2}\,\omega^{(rs)}_{\pm j_1j_2} \cr
&&\pm \frac{15}{8}\,{}^{\star}\HH^j{}_{[i_1\dots i_4}\,\omega^{(rs)}_{\pm i]j}~.
\label{ads2tcfh6}
\eea
Clearly, the conditions on the form bilinears have been arranged as a TCFH as defined in (\ref{tcfhg}) with connection  ${\cal D}^{\cal (\pm)F}$. In fact, the TCFH above has been given in terms of the minimal connection, see  \cite{tcfhgp}. A  consequence of the TCFH above  is that the form bilinears satisfy a generalisation of the CKY equation with respect to ${\cal D}^{\cal (\pm)F}$.

To investigate the (reduced) holonomy of the minimal TCFH connection ${\cal D}^{\cal (\pm)F}$ notice that the TCFH factorises into two parts. One part is spanned
by the form bilinears symmetric in the exchange of $\eta_\pm^r$ and $\eta_\pm^s$ spinors and the other part is spanned by the form bilinears which are skew-symmetric in the exchange of $\eta_\pm^r$ and $\eta_\pm^s$ spinors. Furthermore, ${\cal D}^{\cal (\pm)F}$ acts trivially on the scalars $\rho$ while it acts as a $U(1)$ connection on the scalars $\tilde \rho$. A consequence of this is that the (reduced) holonomy factorises and it is included in (the connected to the identity component of) $U(1)\times GL(133)\times GL(119)$.  Note that the rank of the bundle of symmetric and skew-symmetric form bilinears  in the exchange of $\eta_\pm^r$ and $\eta_\pm^s$ is 136 and 120, respectively. One can also consider the holonomy of the maximal TCFH connection, see \cite{tcfhgp}.  As this acts non-trivially on the scalars, its reduced holonomy is included in (the connected component of) $GL(136)\times GL(120)$.

The factorisation of the holonomy of the  TCFH connections can be also seen from the decomposition of a product of spinor representations of $\mathfrak{spin}(8)$ in terms of forms.  Each $\eta_\pm^r$ spinor can be viewed as a complex chiral $\mathfrak{spin}(8)$ spinor.  The product of two  complex chiral representations, $\Delta^{\pm}_8(\bC)$, of $\mathfrak{spin}(8)$ decomposes as
\bea\label{tensor_product_ads2}
\otimes^2  \Delta_{8}^{\pm}(\bC)=
\Lambda^0(\bC^{8})\oplus \Lambda^2(\bC^{8})\oplus \Lambda^{4\pm}(\bC^{8})~,
\eea
in terms of form representations, where $\Lambda^{4+}(\bC^{8})$ ($\Lambda^{4-}(\bC^{8})$) is the space of the (anti-) self-dual 4-forms on $\bC^{8}$. Then notice that the dimension over the real numbers of the symmetric product, $S^2(\Delta^{\pm}_{8}({\bC}))$,  and skew-symmetric product, $\Lambda^2(\Delta^{\pm}_{8}({\bC}))$,  of two $\Delta^{\pm}_8(\bC)$ representations  is 136 and 120, respectively.  This is exactly the rank of the bundle of the symmetric and skew-symmetric form bilinears in the exchange of $\eta_\pm^r$ and $\eta_\pm^s$ spinors we have considered in the computation of holonomy of TCFH connections. The right-hand-side
of (\ref{tensor_product_ads2}) spans all form bilinears.

The description of the  holonomy of the TCFH connections we have presented above applies to generic backgrounds. As we shall see later for special backgrounds, where some of the form field strengths vanish, the holonomy of the TCFH connections reduces further.

\section{The TCFH of warped AdS$_3$ backgrounds}

\subsection{Fields and Killing spinors}

The fields\footnote{We have not mentioned the $U(1)$-twisted 1-form field strength $P$ of IIB scalars, $P=\xi$, with $\xi$ a $U(1)$-twisted 1-form on the internal space. This is done  to avoid repetition. This equation will also be omitted from the expression of the fields of all AdS backgrounds below. Though it is understood that for the complete description of the fields, it has to be included.} of a warped AdS$_3$ background, AdS$_3\times_w N^7$, can be expressed as
\begin{gather}
g = 2\,\ee^+\ee^- + (\ee^z)^2  + g(N^7)~, \nn
F = \ee^+\wedge \ee^-\wedge \ee^z \wedge Y - {}^{\star_7}Y~, \quad G =  \Phi  \ee^+\wedge \ee^- \wedge \ee^z   + H
~,
\label{metric_ads3}
\end{gather}
where $g(N^7)$ is the internal space metric, $Y$ is a 2-form on $N^7$, and $\Phi$ and $H$ are a $U(1)$-twisted 0- and 3-form on $N^7$, respectively. Furthermore, the pseudo-orhonormal frame can be written as
\bea \label{frame}
\ee^+ = du~, \quad \ee^- = dr-2r(\ell^{-1}\,dz + A^{-1}\dd A)~,\quad \ee^z=A dz~, \quad \ee^i = e^i_I\, dy^I~,
\eea
where  $y^I$  are coordinates of the internal space $N^7$, $(u,r,z)$ are the remaining coordinates of the spacetime, $\ee^i$ is an orthonormal frame on $N^7$, $g(N^7)=\delta_{ij} \ee^i \ee^j$,  and $A$ is the warp factor. It can be seen, after a coordinate transformation, that the spacetime metric $g$ takes the standard warped spacetime form $g=A^2 g_\ell(AdS_3)+ g(N^7)$, where  $g_\ell(AdS_3)$ is the standard metric on AdS$_3$ with radius $\ell$.

The KSEs of warped AdS$_3$ backgrounds can be intergraded over the coordinates $(u,r,z)$, see \cite{sbjggp}, and the Killing spinors can be schematically  expressed as
$\epsilon=\epsilon(u,r,z, \sigma_\pm, \tau_\pm)$, where $\sigma_\pm$ and $\tau_\pm$ depend only on the coordinates of $N^7$ and $\Gamma_\pm\sigma_\pm=\Gamma_\pm\tau_\pm=0$. The integration over the coordinate $z$ introduces a new algebraic KSE on $\sigma_\pm$ and $\tau_\pm$ which will not be explored here but it is essential for the correct counting of Killing spinors of a solution.  This algebraic KSE is in addition to the dilatino KSE of the theory.

A consequence of the gravitino KSE on $\epsilon$ is  that
\bea \label{sigma_tau_KSE}
\nabla^{(\pm)}_i\sigma_{\pm} = 0~, \quad \nabla^{(\pm)}_i\tau_{\pm} = 0~,
\eea
where
\begin{align}\label{KSE3}
\nabla_{i}^{(\pm)} \equiv \nabla_i &\pm \frac{1}{2}\,\partial_i \log A  -\frac{i}{2}\,Q_i \pm \frac{i}{4}\, (\Gamma\slashed{Y})_i\,\Gamma_z \mp \frac{i}{2}\,\slashed{Y}_i\,\Gamma_z  \nn
&+\left( -\frac{1}{96}\,(\Gamma\slashed{H})_i +\frac{3}{32}\,\slashed{H}_i \mp \frac{1}{16}\, \Phi\Gamma_{zi} \right)C*~,
\end{align}
 $\nabla$ is induced on the spinor bundle by the Levi-Civita connection of $g(N^7)$ and $Q$ is a $U(1)$ connection on $N^7$ constructed  from the IIB scalars. The definition of the Clifford algebra operation $C*$  can be found in  section \ref{s1sub1}.

\subsection{The TCFH and holonomy}\label{TCFH_ads3}

Before we proceed to describe the TCFH of the supecovariant connections (\ref{KSE3}), let us first simplify somewhat the analysis. The TCFHs of the form bilinears
constructed using the pairs $(\eta_+^r, \eta_+^s)$ of Killing spinors  are identical, where $\eta_\pm$ stands for either $\sigma_\pm$ or $\eta_\pm$. The reason is that $\sigma_+$ and $\tau_+$ satisfy the same gravitino KSE, see (\ref{sigma_tau_KSE}).  As the bilinears along $N^7$ constructed from $\eta_\pm^r$ and $\eta_\mp^s$ vanish, it remains to consider the TCFH constructed from the bilinears of $\eta_-$.  This TCFH can be easily deduced from that of the $\eta_+$ form
bilinears after appropriately compensating for the differences in the signs of some of the  terms in the supercovariant derivatives
$\nabla^{(+)}$ and $\nabla^{(-)}$, see (\ref{KSE3}).  There is also an additional sign required in all terms that contain a Hodge duality operation on the fluxes  that appear in the TCFHs. This is  a consequence of conditions $\Gamma_\pm \eta_\pm=0$ on the spinors, see also below.

A consequence of the discussion above is that, without loss of generality, we can focus on the TCFH associated with the bilinears of $\sigma_+$ Killing spinors. Setting $\sigma_+=\sigma$, one finds that a basis in the space of form bilinears on $N^7$ is
\begin{gather}
\rho^{rs} = \langle \sigma^r, \sigma^s \rangle~, \quad \tilde{\rho}^{rs} = \langle \sigma^r, C\bar{\sigma}^s \rangle~, \nn
\kappa^{rs} = \langle \sigma^r, \Gamma_z \Gamma_{i}\, \sigma^s \rangle \, \ee^i~, \quad  \tilde{\kappa}^{rs} = \langle \sigma^r, \Gamma_z \Gamma_{i}\, C\bar{\sigma}^s \rangle \, \ee^i ~, \nn
\omega^{rs} = \frac{1}{2}\, \langle \sigma^r, \Gamma_{i_1 i_2}\, \sigma^s \rangle\, \ee^{i_1}\wedge \ee^{i_2}~, \quad \tilde{\omega}^{rs} = \frac{1}{2}\, \langle \sigma^r, \Gamma_{i_1 i_2}\, C\bar{\sigma}^s \rangle\, \ee^{i_1}\wedge \ee^{i_2}~, \nn
\psi^{rs} = \frac{1}{3!}\, \langle \sigma^r, \Gamma_z \Gamma_{i_1i_2i_3}\, \sigma^s \rangle \, \ee^{i_1}\wedge \ee^{i_2}\wedge \ee^{i_3}~, \quad \tilde{\psi}^{rs} = \frac{1}{3!}\, \langle \sigma^r, \Gamma_z \Gamma_{i_1i_2i_3}\, C\bar{\sigma}^s \rangle \, \ee^{i_1}\wedge \ee^{i_2}\wedge \ee^{i_3}~.
\label{ads3_bilinears}
\end{gather}
It turns out that $\tilde{\rho}^{rs}$, $\tilde{\psi}^{rs}$, $\rep \rho^{rs}$, $\imp\kappa^{rs}$, $\rep\psi^{rs}$ and $\imp\omega^{rs}$ are symmetric, while $\tilde{\kappa}^{rs}$, $\tilde{\omega}^{rs}$, $\rep\kappa^{rs}$, $\imp\psi^{rs}$, $\rep\omega^{rs}$ and $\imp \rho^{rs}$ are skew-symmetric in the exchange of the spinors $\sigma^r$ and $\sigma^s$. Note that as a consequence of the IIB chirality of spinors $\sigma_\pm$ and the condition $\Gamma_\pm \sigma_\pm=0$, one has that $\Gamma_{(7)}\Gamma_z\,\sigma_{\pm}=\pm \sigma_{\pm}$, where $\Gamma_{(7)} =\prod_{i=1}^7 \Gamma_i $.  This justifies the choice of the above basis in the space of form bilinears up to a Hodge duality operation on $N^7$.  As it has already been mentioned in the beginning of the section,  the sign of the condition  $\Gamma_{(7)}\Gamma_z\,\sigma_{\pm}=\pm \sigma_{\pm}$ accounts for the additional sign required in the terms that contain a Hodge duality operation on the fluxes in the TCFH   associated with the $\sigma_+$ form bilinears relative to  the same terms of the TCFH constructed from the $\sigma_-$ form bilinears.

The computation of the TCFH for the bilinears (\ref{ads3_bilinears}) is similar to that described for warped AdS$_2$ backgrounds in the previous section. After some computation, one finds that
\bea
{\cal D}^{\cal (+)F}_i \rho^{rs} &\defeq& \nabla_i\, \rho^{rs} = -\partial_i \log A\, \rho^{rs} - i\, Y_i{}^j\,\kappa^{rs}_j + \frac{1}{48}\, \rep \{{}^{\star}H_i{}^{j_1j_2j_3}\,\tilde{\psi}^{rs}_{j_1j_2j_3}\}  \cr
&&-\frac{3i}{16}\,\imp\{H_i{}^{j_1j_2}\,\tilde{\omega}^{rs}_{j_1j_2} \} +\frac{i}{8}\,\imp\{\Phi \tilde{\kappa}^{rs}_i\} ~,
\label{ads3tcfh1}
\eea
\bea
{\cal D}^{\cal (+)F}_i \kappa^{rs}_k &\defeq& \nabla_i\,\kappa^{rs}_k + \partial_i \log A\,\kappa^{rs}_k + \frac{i}{4}\, \imp\{H_i{}^{j_1j_2}\,\tilde{\psi}^{rs}_{kj_1j_2}\}  \cr
&&= -\frac{i}{6}\,{}^{\star}Y_{ik}{}^{j_1j_2j_3}\psi^{rs}_{j_1j_2j_3} + i\,Y_{ik}\,\rho^{rs} + \frac{i}{48}\,\delta_{ik}\,\imp \{H^{j_1j_2j_3}\tilde{\psi}^{rs}_{j_1j_2j_3}\} \cr
&& +\frac{i}{8}\,\imp\{H^{j_1j_2}{}_{[i}\,\tilde{\psi}^{rs}_{k]j_1j_2}\}-\frac{i}{8}\,\delta_{ik}\,\imp\{ \Phi\,\tilde{\rho}^{rs} \} -\frac{1}{16}\,\rep\{ {}^{\star}H_{ik}{}^{j_1j_2}\,\tilde{\omega}^{rs}_{j_1j_2}\}
 \cr
 &&
 -\frac{3}{8}\,\rep\{H_{ik}{}^{j}\,\tilde{\kappa}^{rs}_j\} +\frac{1}{8}\,\rep\{\Phi \,\tilde{\omega}^{rs}_{ik}\} ~,
\label{ads3tcfh2}
\eea
\bea
{\cal D}^{\cal (+)F}_i \omega^{rs}_{i_1i_2} &\defeq& \nabla_i\,\omega^{rs}_{i_1i_2} + \partial_i\log A\, \omega^{rs}_{i_1i_2} + 2i\,Y_{i}{}^{j}\,\psi^{rs}_{i_1i_2j} + \frac{i}{2}\,\imp\{{}^{\star}H^{j_1j_2}{}_{i[i_1}\tilde{\psi}^{rs}_{i_2]j_1j_2}\} -\rep\{H^j{}_{i[i_1}\,\tilde{\omega}^{rs}_{i_2]j}\} \cr
&&=-i\,Y^{j_1j_2}\,\delta_{i[i_1}\psi^{rs}_{i_2]j_1j_2} - 3i\,Y^j{}_{[i}\psi^{rs}_{i_1i_2]j} +\frac{i}{8}\,\delta_{i[i_1}\,\imp\{{}^{\star}H_{i_2]}{}^{j_1j_2j_3}\tilde{\psi}^{rs}_{j_1j_2j_3}\} \cr
&&+\frac{9i}{16}\,\imp\{{}^{\star}H^{j_1j_2}{}_{[i_1i_2}\tilde{\psi}^{rs}_{i]j_1j_2}\} + \frac{3i}{8}\,\imp\{H_{ii_1i_2}\tilde{\rho}^{rs}\} + \frac{i}{8}\,\imp\{\Phi\,\tilde{\psi}^{rs}_{ii_1i_2}\} \cr
&& -\frac{1}{8}\,\delta_{i[i_1}\,\rep\{H_{i_2]}{}^{j_1j_2}\,\tilde{\omega}^{rs}_{j_1j_2}\} -\frac{3}{8}\,\rep\{H^j{}_{[i_1i_2}\,\tilde{\omega}^{rs}_{i]j}\} -\frac{1}{8}\,\rep\{{}^{\star}H_{ii_1i_2}{}^j\,\tilde{\kappa}^{rs}_j\} \cr
&&-\frac{1}{4}\,\delta_{i[i_1}\,\rep\{\Phi\,\tilde{\kappa}^{rs}_{i_2]}\}
\label{ads3tcfh3}
\eea
\bea
{\cal D}^{\cal (+)F}_i \psi^{rs}_{i_1i_2i_3} &\defeq&  \nabla_i\,\psi^{rs}_{i_1i_2i_3} + \partial_i \log A\, \psi^{rs}_{i_1i_2i_3} - 6i\, Y_{i[i_1}\,\omega^{rs}_{i_2i_3]} + \frac{3}{2}\, \rep\{H^j{}_{i[i_1}\,\tilde{\psi}^{rs}_{i_2i_3]j}\} \cr
&& -\frac{3i}{2}\,\imp\{H_{i[i_1i_2}\,\tilde{\kappa}^{rs}_{i_3]}\} +\frac{3i}{2}\,\imp\{{}^{\star}H^j{}_{i[i_1i_2}\,\tilde{\omega}^{rs}_{i_3]j}\} \cr
&&= -i\,{}^{\star}Y_{i_1i_2i_3i}{}^j\,\kappa^{rs}_j + 6i\,\delta_{i[i_1}\,Y_{i_2}{}^j\,\omega^{rs}_{i_3]j} - 6i\,Y_{[ii_1}\,\omega^{rs}_{i_2i_3]} - \frac{3}{8}\,\delta_{i[i_1}\,\rep\{H_{i_2}{}^{j_1j_2}\,\tilde{\psi}^{rs}_{i_3]j_1j_2}\} \cr
&&+\frac{3}{4}\,\rep\{H^j{}_{[ii_1}\,\tilde{\psi}^{rs}_{i_2i_3]j}\} + \frac{1}{8}\,\rep\{{}^{\star}H_{ii_1i_2i_3}\,\tilde{\rho}^{rs}\} - \frac{1}{48}\,\rep\{{}^{\star}\Phi_{ii_1i_2i_3}{}^{j_1j_2j_3}\,\tilde{\psi}^{rs}_{j_1j_2j_3}\} \cr
&& -\frac{3i}{8}\,\delta_{i[i_1}\,\imp\{H_{i_2i_3]}{}^j\,\tilde{\kappa}^{rs}_j\} + \frac{i}{2}\,\imp\{H_{[i_1i_2i_3}\,\tilde{\kappa}^{rs}_{i]}\} -\frac{9i}{16}\,\delta_{i[i_1}\,\imp\{{}^{\star}H_{i_2i_3]}{}^{j_1j_2}\,\tilde{\omega}^{rs}_{j_1j_2}\} \cr
&& -\frac{3i}{2}\,\imp\{{}^{\star}H^j{}_{[i_1i_2i_3}\,\tilde{\omega}^{rs}_{i]j}\} -\frac{3i}{8}\,\delta_{i[i_1}\,\imp\{\Phi\,\tilde{\omega}^{rs}_{i_2i_3]}\} ~,
\label{ads3tcfh4}
\eea
\bea
{\cal D}^{\cal (+)F}_i\tilde{\rho}^{rs} &\defeq& \nabla_i\,\tilde{\rho}^{rs} +(iQ_i+\partial_i\log A)\,\tilde{\rho}^{rs} =- \frac{i}{2}\,Y^{j_1j_2}\,\tilde{\psi}^{rs}_{ij_1j_2} + \frac{1}{48}\,{}^{\star}\HH_i{}^{j_1j_2j_3}\,\psi^{(rs)}_{j_1j_2j_3} \cr
&&-\frac{3}{16}\,\HH_i{}^{j_1j_2}\,\omega^{(rs)}_{j_1j_2} + \frac{1}{8}\,\PP\kappa^{(rs)}_i~,
\label{ads3tcfh5}
\eea
\bea
{\cal D}^{\cal (+)F}_i\tilde{\kappa}^{rs}_k &\defeq& \nabla_i\,\tilde{\kappa}^{rs}_k + (iQ_i+\partial_i\log A)\,\tilde{\kappa}^{rs}_k +2i\,Y_i{}^j\,\tilde{\omega}^{rs}_{kj} + \frac{1}{4}\,\HH_i{}^{j_1j_2}\,\psi^{[rs]}_{kj_1j_2} \cr
&&= \frac{i}{2}\,\delta_{ik}\,Y^{j_1j_2}\,\tilde{\omega}^{rs}_{j_1j_2} + 2i\,Y^j{}_{[k}\,\tilde{\omega}^{rs}_{i]j} +\frac{1}{48}\,\delta_{ik}\,\HH^{j_1j_2j_3}\,\psi^{[rs]}_{j_1j_2j_3} + \frac{1}{8}\,\HH^{j_1j_2}{}_{[i}\,\psi^{[rs]}_{k]j_1j_2} \cr
&& -\frac{1}{16}\,{}^{\star}\HH_{ik}{}^{j_1j_2}\,\omega^{[rs]}_{j_1j_2} -\frac{3}{8}\,\HH_{ik}{}^j\,\kappa^{[rs]}_j +\frac{1}{8}\,\PP\,\omega^{[rs]}_{ik} -\frac{1}{8}\,\delta_{ik}\,\PP\, \rho^{[rs]} ~,
\label{ads3tcfh6}
\eea
\bea
{\cal D}^{\cal (+)F}_i\tilde{\omega}^{rs}_{i_1i_2} &\defeq& \nabla_i\,\tilde{\omega}^{rs}_{i_1i_2} + (iQ_i+\partial_i \log A)\,\tilde{\omega}^{rs}_{i_1i_2} + 4i\,Y_{i[i_1}\,\tilde{\kappa}^{rs}_{i_2]} - \HH^j_{i[i_1}\,\omega^{[rs]}_{i_2]j} +\frac{1}{2}\,{}^{\star}\HH^{j_1j_2}{}_{i[i_1}\,\psi^{[rs]}_{i_2]j_1j_2} \cr
&&=-\frac{i}{2}\,{}^{\star}Y_{ii_1i_2}{}^{j_1j_2}\,\tilde{\omega}^{rs}_{j_1j_2} +2i\,\delta_{i[i_1}\,Y_{i_2]}{}^j\,\tilde{\kappa}^{rs}_j + 3i\,Y_{[i_1i_2}\,\tilde{\kappa}^{rs}_{i]} -\frac{1}{8}\,\delta_{i[i_1}\,\HH_{i_2]}{}^{j_1j_2}\,\omega^{[rs]}_{j_1j_2} \cr
&&-\frac{3}{8}\,\HH^j{}_{[i_1i_2}\,\omega^{[rs]}_{i]j} -\frac{1}{8}\,{}^{\star}\HH_{ii_1i_2}{}^j\,\kappa^{[rs]}_j +\frac{1}{8}\,\delta_{i[i_1}\,{}^{\star}\HH_{i_2]}{}^{j_1j_2j_3}\,\psi^{[rs]}_{j_1j_2j_3}+ \frac{9}{16}\,{}^{\star}\HH^{j_1j_2}{}_{[i_1i_2}\,\psi^{[rs]}_{i]j_1j_2} \cr
&&+\frac{3}{8}\,\HH_{ii_1i_2}\,\rho^{[rs]} + \frac{1}{8}\,\PP\,\psi^{[rs]}_{ii_1i_2} -\frac{1}{4}\,\PP\, \delta_{i[i_1}\,\kappa^{[rs]}_{i_2]} ~,
\label{ads3tcfh7}
\eea
\bea
{\cal D}^{\cal (+)F}_i\tilde{\psi}^{rs}_{i_1i_2i_3} &\defeq& \nabla_i\,\tilde{\psi}^{rs}_{i_1i_2i_3} + (iQ_i +\partial_i\log A)\,\tilde{\psi}^{rs}_{i_1i_2i_3} + 3i\,{}^{\star}Y^{j_1j_2}{}_{i[i_1i_2}\, \tilde{\psi}^{rs}_{i_3]j_1j_2}-\frac{3}{2}\,\HH_{i[i_1i_2}\,\kappa^{(rs)}_{i_3]} \cr
&& + \frac{3}{2}\,\HH^j{}_{i[i_1}\,\psi^{(rs)}_{i_2i_3]j} + \frac{3}{2}\,{}^{\star}\HH^j{}_{i[i_1i_2}\,\omega^{(rs)}_{i_3]j} \cr
&&= -3i\,\delta_{i[i_1}\,Y_{i_2i_3]}\tilde{\rho}^{rs} +\frac{i}{2} \,\delta_{i[i_1}\,{}^{\star}Y_{i_2i_3]}{}^{j_1j_2j_3}\tilde{\psi}^{rs}_{j_1j_2j_3} -2i\,{}^{\star}Y^{j_1j_2}{}_{[i_1i_2i_3}\tilde{\psi}^{rs}_{i]j_1j_2} \cr
&&-\frac{3}{8}\,\delta_{i[i_1}\,\HH_{i_2i_3]}{}^j\,\kappa^{(rs)}_j +\frac{1}{2}\,\HH_{[i_1i_2i_3}\,\kappa^{(rs)}_{i]} -\frac{3}{8}\,\delta_{i[i_1}\,\HH_{i_2}{}^{j_1j_2}\,\psi^{(rs)}_{i_3]j_1j_2} -\frac{3}{4}\,\HH^j{}_{[i_1i_2}\,\psi^{(rs)}_{i_3i]j}\cr
&&+ \frac{1}{8}\,{}^{\star}\HH_{ii_1i_2i_3}\, \rho^{(rs)} - \frac{9}{16}\,\delta_{i[i_1}{}^{\star}\HH_{i_2i_3]}{}^{j_1j_2}\,\omega^{(rs)}_{j_1j_2} -\frac{3}{2}{}^{\star}\HH^j{}_{[i_1i_2i_3}\,\omega^{(rs)}_{i]j} - \frac{3}{8}\,\PP\,\delta_{i[i_1}\omega^{(rs)}_{i_2i_3]} \cr
&& -\frac{1}{48}\,{}^{\star}\PP_{ii_1i_2i_3}{}^{j_1j_2j_3}\,\psi^{(rs)}_{j_1j_2j_3}  ~.
\label{ads3tcfh8}
\eea

The TCFH above has been expressed in terms of the minimal connection ${\cal D}^{\cal (+)F}$. As for the AdS$_2$ case, to find the holonomy of this connection for generic backgrounds observe
that it preserves the domain of symmetric and skew-symmetric form bilinears in the exchange of the spinors $\sigma^r$ and $\sigma^s$. Furthermore, it acts trivially on the scalars $\rho^{rs}$, as a $U(1)$ connection on the scalars $\tilde \rho^{rs}$ and with the Levi-Civita connection on the 1-form bilinear $A \rep \kappa^{rs}$. Therefore, the (reduced) holonomy of the minimal connection is included in (the connected component of) $U(1)\times GL(133)\times SO(7)\times GL(112)$, where the  $U(1)\times GL(133)$ subgroup is associated with the symmetric form bilinears while the  rest is associated with the skew-symmetric ones. The holonomy of the maximal TCFH connection is expected to be included in $GL(136)\times GL(120)$ as its action on all form bilinears is not trivial though it still preserves the subspaces of symmetric and skew-symmetric form bilinears.
Similar conclusions hold for the connections of the TCFHs of the rest of the form bilinears constructed from the spinors $\sigma_\pm$ and $\tau_\pm$.

\section{The TCFH of warped AdS$_4$ backgrounds}

\subsection{Fields and spinors}

The fields of warped AdS$_4$ backgrounds, AdS$_4\times_w N^6$, can be written as
\begin{gather}
g= 2\,\ee^+\ee^- + (\ee^z)^2 +(\ee^x)^2  + g(N^6)~, \nn
F = \ee^+\wedge \ee^-\wedge \ee^z \wedge \ee^x \wedge Y + {}^{\star_6}Y~, \quad G = H
~,
\end{gather}
where $g(N^6)$ is the metric on the internal space $N^6$, and $Y$ and $H$ are a 1-form  and  a U(1)-twisted 3-form on $N^6$, respectively. Furthermore, the components $(\ee^+, \ee^-, \ee^z, \ee^i)$ of pseudo-orthonormal frame are defined as for the AdS$_3$ backgrounds in (\ref{frame}) with the understanding that the warp factor $A$ is a function on $N^6$ and $\ee^i$ is an orthonormal frame on $N^6$,  $g(N^6)=\delta_{ij} \ee^i \ee^j$, where $y^I$ are coordinates of $N^6$ and $(u,r,z,x)$ are the remaining coordinates of the spacetime.  Moreover, the remaining component of the pseudo-orthonormal frame is $\ee^x= A e^{z/\ell} dx$. It can be seen after a coordinate transformation that the spacetime metric takes the standard warped form $g=A^2 g_\ell(AdS_4)+ g(N^6)$, where $g_\ell(AdS_4)$ is the standard metric on AdS$_4$ with radius $\ell$.

The IIB KSEs for warped AdS$_4$ backgrounds have been solved in \cite{sbjggp}. Integrating the KSEs over the coordinates $(u,r,z, x)$, the Killing spinors $\epsilon$ can be expressed as $\epsilon=\epsilon(u,r,z,x, \sigma_\pm, \tau_\pm)$, where the spinors\footnote{Unlike for the AdS$_3$ backgrounds that $\sigma_\pm$ and $\tau_\pm$ are unrelated, the $\sigma_\pm$ and $\tau_\pm$ spinors for all warped AdS$_k$ backgrounds, $k>3$, are related with certain Clifford algebra operations \cite{sbjggp}.} $\sigma_\pm$ and $\tau_\pm$ depend only on the coordinates of $N^6$ and satisfy $\Gamma_\pm\sigma_\pm=\Gamma_\pm\tau_\pm=0$.  Furthermore, the gravitino KSE implies that
$ \nabla^{(\pm)}_i\sigma_{\pm} = 0$ and $\nabla^{(\pm)}_i\tau_{\pm} = 0$, where the supercovariant derivatives are
\begin{align}\label{KSE4}
\nabla_{i}^{(\pm)} \equiv \nabla_i &\pm \frac{1}{2}\,\partial_i \log A  -\frac{i}{2}\,Q_i \mp \frac{i}{2}\, (\Gamma\slashed{Y})_i\,\Gamma_{xz} \pm \frac{i}{2}\, Y_i\,\Gamma_{xz}  \nn
&+\left( -\frac{1}{96}\,(\Gamma\slashed{H})_i +\frac{3}{32}\,\slashed{H}_i \right)C*~,
\end{align}
 and the Clifford algebra operation $C$ is defined as in the AdS$_2$ case.

\subsection{The TCFH and holonomy}

As for warped  AdS$_3$ backgrounds, it suffices to describe only the TCFH of  $\sigma_+$ spinor form bilinears. The TCFH of the form bilinears of all other spinors can be derived from that of the $\sigma_+$ spinors. The method of this derivation  has already been described in the AdS$_3$ case. In addition, the TCFH of warped AdS$_4$ backgrounds factorises on the subspaces of even- and odd-degree (twisted) forms on the internal space $N^6$.  Because of this the two cases will be treated separately.
A basis in the space of even-degree form bilinears of $\sigma=\sigma_+$ spinors can be chosen as
\begin{gather}
\rho^{rs} = \langle \sigma^r, \sigma^s \rangle~, \quad \tilde{\rho}^{rs} = \langle \sigma^r, C\bar{\sigma}^s \rangle~, \nn
\rrho^{rs} = \langle \sigma^r, \Gamma_{xz}\, \sigma^s \rangle~, \quad \trrho^{rs} = \langle \sigma^r, \Gamma_{xz}\, C\bar{\sigma}^s \rangle ~, \nn
\omega^{rs} = \frac{1}{2}\, \langle \sigma^r, \Gamma_{i_1 i_2}\, \sigma^s \rangle\, \ee^{i_1}\wedge \ee^{i_2}~, \quad \tilde{\omega}^{rs} = \frac{1}{2}\, \langle \sigma^r, \Gamma_{i_1 i_2}\, C\bar{\sigma}^s \rangle\, \ee^{i_1}\wedge \ee^{i_2} ~, \nn
\romega^{rs} = \frac{1}{2}\, \langle \sigma^r, \Gamma_{xz}\Gamma_{i_1 i_2}\, \sigma^s \rangle\, \ee^{i_1}\wedge \ee^{i_2}~, \quad \tromega^{rs} = \frac{1}{2}\, \langle \sigma^r, \Gamma_{xz} \Gamma_{i_1 i_2}\, C\bar{\sigma}^s \rangle\, \ee^{i_1}\wedge \ee^{i_2}~.
\label{ads4_bilinears}
\end{gather}
It turns out  that $\tilde{\rho}^{rs}, \tromega^{rs}, \rep \rho^{rs}, \imp\rrho^{rs}, \imp\omega^{rs}$ and $\rep \romega^{rs}$ are symmetric while $\trrho^{rs}, \tilde{\omega}^{rs}$, $\imp \rho^{rs}$, $\rep\rrho^{rs}$, $\rep \omega^{rs}$ and $\imp \romega^{rs}$ are skew-symmetric in the exchange of spinors $\sigma^r$ and $\sigma^s$.

A direct computation reveals that the TCFH expressed in terms of the minimal connection ${\cal D}^{\cal F}$  is
\bea
{\cal D}^{\cal F}_i \rho^{rs} &\defeq& \nabla_i\, \rho^{rs}  =-\partial_i \log A\, \rho^{rs} - i\,Y_i\,\rrho^{rs} - \frac{1}{16}\, \rep \{ {}^{\star}H_{i}{}^{j_1j_2}\,\tromega^{rs}_{j_1j_2} \} \cr
&&-\frac{3i}{16}\,\imp\{H_{i}{}^{j_1j_2}\,\tilde{\omega}^{rs}_{j_1j_2}\} ~,
\label{ads4tcfh1}
\eea
\bea
{\cal D}^{\cal F}_i \rrho^{rs} &\defeq& \nabla_i\, \rrho^{rs} = -\partial_i \log A\, \rrho^{rs} + i\, Y_i\, \rho^{rs} -\frac{3i}{16}\, \imp \{ H_{i}{}^{j_1j_2}\, \tromega^{rs}_{j_1j_2} \} \cr
&&+ \frac{1}{16}\,\rep\{{}^{\star}H_{i}{}^{j_1j_2}\,\tilde{\omega}^{rs}_{j_1j_2}\} ~,
\label{ads4tcfh2}
\eea
\bea
{\cal D}^{\cal F}_i \omega^{rs}_{i_1i_2} &\defeq& \nabla_i\, \omega^{rs}_{i_1i_2} + \partial_i\log A\, \omega^{rs}_{i_1i_2} + 2i\, Y_i\, \romega^{rs}_{i_1i_2} - i\,\imp\{ {}^{\star}H^j{}_{i[i_1}\,\tromega^{rs}_{i_2]j} \} -\rep\{H^j{}_{i[i_1}\,\tilde{\omega}^{rs}_{i_2]j}\} \cr
&&= -2i\, Y^j\, \delta_{i[i_1}\,\romega^{rs}_{i_2]j} + 3i\, Y_{[i}\,\romega^{rs}_{i_1i_2]} - \frac{3i}{8}\,\delta_{i[i_1}\imp \{ {}^{\star}H_{i_2]}{}^{j_1j_2}\,\tromega^{rs}_{j_1j_2} \} - \frac{9i}{8}\, \imp \{ {}^{\star} H^j{}_{[i_1i_2}\,\tromega^{rs}_{i]j} \} \cr
&&+ \frac{3i}{8}\, \imp\{ H_{ii_1i_2}\tilde{\rho}^{rs} \} -\frac{1}{8}\,\delta_{i[i_1}\,\rep\{H_{i_2]}{}^{j_1j_2}\,\tilde{\omega}^{rs}_{j_1j_2}\} - \frac{3}{8}\,\rep\{H^j{}_{[i_1i_2}\,\tilde{\omega}^{rs}_{i]j}\} \cr
&&+\frac{1}{8}\,\rep\{{}^{\star}H_{ii_1i_2}\,\trrho^{rs}\} ~,
\label{ads4tcfh3}
\eea
\bea
{\cal D}^{\cal F}_i \romega^{rs}_{i_1i_2} &\defeq& \nabla_i\, \romega^{rs}_{i_1i_2} + \partial_i \log A \, \romega^{rs}_{i_1i_2} -2i\, Y_i\,\omega^{rs}_{i_1i_2} - \rep \{ H^j{}_{i[i_1}\tromega^{rs}_{i_2]j} \} +i\imp\{{}^{\star}H^j{}_{i[i_1}\,\tilde{\omega}^{rs}_{i_2]j}\} \cr
&&= 2i\, Y^j\, \delta_{i[i_1}\omega^{rs}_{i_2]j} - 3i\, Y_{[i}\,\omega^{rs}_{i_1i_2]} - \frac{1}{8}\,\rep \{ {}^{\star}H_{ii_1i_2}\,\tilde{\rho}^{rs} \} - \frac{1}{8}\,\delta_{i[i_1}\,\rep \{ H_{i_2]}{}^{j_1j_2}\,\tromega^{rs}_{j_1j_2} \} \cr
&&- \frac{3}{8}\,\rep \{ H^j{}_{[ii_1}\,\romega^{rs}_{i_2]j} \} +\frac{3i}{8}\,\delta_{i[i_1}\,\imp\{{}^{\star}H_{i_2]}{}^{j_1j_2}\,\tilde{\omega}^{rs}_{j_1j_2} \} +\frac{9i}{8}\,\imp\{{}^{\star}H^j{}_{[i_1i_2}\,\tilde{\omega}^{rs}_{i]j} \} \cr
&&+\frac{3i}{8}\,\imp\{H_{ii_1i_2}\trrho^{rs} \} ~,
\label{ads4tcfh4}
\eea
\bea
{\cal D}^{\cal F}_i \tilde{\rho}^{rs} &\defeq& \nabla_i\, \tilde{\rho}^{rs} +(iQ_i +\partial_i \log A)\,\tilde{\rho}^{rs}= - i\, Y^j\, \romega^{(rs)}_{ij} - \frac{1}{16}\, {}^{\star}\HH_i{}^{j_1j_2}\,\romega^{(rs)}_{j_1j_2} \cr
&&-\frac{3}{16}\, \HH_i{}^{j_1j_2}\,\omega^{(rs)}_{j_1j_2}~,
\label{ads4tcfh5}
\eea
\bea
{\cal D}^{\cal F}_i \trrho^{rs} &\defeq& \nabla_i\,\trrho^{rs}+(iQ_i+\partial_i\log A)\,\trrho^{rs} = i\,Y^j\,\tilde{\omega}^{rs}_{ij} +\frac{1}{16}\,{}^{\star}\HH_i{}^{j_1j_2}\,\omega^{[rs]}_{j_1j_2} \cr
&&-\frac{3}{16}\,\HH_i{}^{j_1j_2}\,\romega^{[rs]}_{j_1j_2}
\label{ads4tcfh6}
\eea
\bea
{\cal D}^{\cal F}_i\tilde{\omega}^{rs}_{i_1i_2} &\defeq& \nabla_i\,\tilde{\omega}^{rs}_{i_1i_2} + (iQ_i+\partial_i\log A)\,\tilde{\omega}^{rs}_{i_1i_2} - \HH^j{}_{i[i_1}\,\omega^{[rs]}_{i_2]j} - \,{}^{\star}\HH^j{}_{i[i_1}\,\romega^{[rs]}_{i_2]j} \cr
&&= -\frac{i}{2}\,{}^{\star}Y_{ii_1i_2}{}^{j_1j_2}\,\tilde{\omega}^{rs}_{j_1j_2} + 2i\,\delta_{i[i_1}\,Y_{i_2]}\,\trrho^{rs} - \frac{1}{8}\,\delta_{i[i_1}\,\HH_{i_2]}{}^{j_1j_2}\,\omega^{[rs]}_{j_1j_2} \cr
&& -\frac{3}{8}\,\HH^j{}_{[i_1i_2}\,\omega^{[rs]}_{i]j}  + \frac{1}{8}\,{}^{\star}\HH_{ii_1i_2}\,\rrho^{[rs]} - \frac{3}{8}\,\delta_{i[i_1}\,\HH_{i_2]}{}^{j_1j_2}\,\romega^{[rs]}_{j_1j_2} \cr
&&-\frac{9}{8}\,{}^{\star}\HH^j{}_{[i_1i_2}\,\romega^{[rs]}_{i]j} + \frac{3}{8}\,\HH_{ii_1i_2}\,\rho^{[rs]} ~,
\label{ads4tcfh7}
\eea
\bea
{\cal D}^{\cal F}_i \tromega^{rs}_{i_1i_2} &\defeq& \nabla_i\,\tromega^{rs}_{i_1i_2} + (iQ_i + \partial_i \log A)\,\tromega^{rs}_{i_1i_2} + {}^{\star} \HH^j{}_{i[i_1}\,\omega^{(rs)}_{i_2]j} - \HH^j{}_{i[i_1}\,\romega^{(rs)}_{i_2]j} \cr
&&= -\frac{i}{2}\, {}^{\star} Y_{ii_1i_2}{}^{j_1j_2}\,\tromega^{rs}_{j_1j_2} -2i\, \delta_{i[i_1}\, Y_{i_2]}\, \tilde{\rho}^{rs}- \frac{1}{8}\, {}^{\star}\HH_{ii_1i_2}\,\rho^{(rs)} -\frac{1}{8}\, \delta_{i[i_1}\, \HH_{i_2]}{}^{j_1j_2}\, \romega^{(rs)}_{j_1j_2} \cr
&&- \frac{3}{8}\, \HH^j{}_{[ii_1}\, \romega^{(rs)}_{i_2]j} + \frac{3}{8}\, \delta_{i[i_1}\,{}^{\star}\HH_{i_2]}{}^{j_1j_2}\,\omega^{(rs)}_{j_1j_2} + \frac{9}{8}\,{}^{\star}\HH^j{}_{[i_1i_2}\,\omega^{(rs)}_{i]j} \cr
&&+ \frac{3}{8}\, \HH_{ii_1i_2}\,\rrho^{(rs)} ~,
\label{ads4tcfh8}
\eea
where we have used that $(\prod_i \Gamma_i) \Gamma_{xz}\sigma_\pm=\pm \sigma_\pm$ which is a consequence of $\Gamma_\pm\sigma_\pm=0$ and the chirality of the IIB spinors.
The (reduced) holonomy of the minimal connection ${\cal D}^{\cal F}$ can be computed as in previous cases yielding that it must be contained  in (the connected component of) $\times^2 (U(1)\times GL(60))$.

Next, a basis in the space of odd-degree form bilinears of $\sigma=\sigma_+$ spinors can be chosen as
\begin{gather}
\kappa^{rs} = \langle \sigma^r, \Gamma_{zi}\sigma^s \rangle\, \ee^i~,  \quad \tilde{\kappa}^{rs} = \langle \sigma^r, \Gamma_{zi} C\bar{\sigma}^s \rangle\, \ee^i~, \nn
\mathring{\kappa}^{rs} = \langle \sigma^r, \Gamma_{xi}\sigma^s \rangle\, \ee^i~,  \quad \mathring{\tilde{\kappa}}^{rs} = \langle \sigma^r~, \Gamma_{xi} C\bar{\sigma}^s \rangle\, \ee^i~, \nn
\psi^{rs} = \frac{1}{3!}\, \langle \sigma^r, \Gamma_{zi_1 i_2i_3 }\, \sigma^s \rangle\, \ee^{i_1}\wedge \ee^{i_2}\wedge \ee^{i_3}~, \quad \tilde{\psi}^{rs} = \frac{1}{3!}\, \langle \sigma^r, \Gamma_{zi_1 i_2 i_3 }\, C\bar{\sigma}^s \rangle\, \ee^{i_1}\wedge \ee^{i_2}\wedge \ee^{i_3} ~. \nn
\label{ads4_bilinears2}
\end{gather}
The associated TCFH is
\bea
{\cal D}^{\cal F}_i\,\kappa^{rs}_k &\defeq& \nabla_i\kappa^{rs}_k + \partial_i\log A \,\kappa^{rs}_k - \frac{i}{4}\,\imp\lbrace {}^{\star}H_i{}^{j_1j_2}\tilde{\psi}^{rs}_{kj_1j_2} \rbrace \cr
&&=-\frac{i}{6} {}^{\star}Y_{ik}{}^{j_1j_2j_3}\psi^{rs}_{j_1j_2j_3} +\frac{i}{48}\,\delta_{ik}\,\imp\lbrace H^{j_1j_2j_3}\,\tilde{\psi}^{rs}_{j_1j_2j_3} \rbrace + \frac{i}{8}\,\imp\lbrace H^{j_1j_2}{}_{[i}\,\tilde{\psi}^{rs}_{k]j_1j_2} \rbrace \cr
&& +\frac{1}{8}\,\rep\lbrace {}^{\star}H_{ik}{}^j\tkkappa^{rs}_j \rbrace - \frac{3}{8}\,\rep\lbrace H_{ik}{}^j\tilde{\kappa}^{rs}_j \rbrace~,
\label{ads4tcfh9}
\eea
\bea
{\cal D}^{\cal F}_i\kkappa^{rs}_k &\defeq& \nabla_i\,\kkappa^{rs}_k + \partial_i\log A \,\kkappa^{rs}_k + \frac{i}{4}\,\imp\lbrace {}^{\star}H_i{}^{j_1j_2}\tilde{\psi}^{rs}_{kj_1j_2} \rbrace \cr
&&=-iY^j\psi^{rs}_{ikj} +\frac{i}{16}\,\delta_{ik}\,\imp\lbrace {}^{\star}H^{j_1j_2j_3}\,\tilde{\psi}^{rs}_{j_1j_2j_3} \rbrace +\frac{3i}{8}\,\imp\lbrace {}^{\star}H^{j_1j_2}{}_{[i}\psi^{rs}_{k]j_1j_2} \rbrace \cr
&&-\frac{1}{8}\,\rep\lbrace {}^{\star}H_{ik}{}^j\tilde{\kappa}^{rs}_j \rbrace - \frac{3}{8}\,\rep\lbrace H_{ik}{}^j\tkkappa^{rs}_j \rbrace~,
\label{ads4tcfh10}
\eea
\bea
{\cal D}^{\cal F}_i\psi^{rs}_{i_1i_2i_3} &\defeq& \nabla_i\,\psi^{rs}_{i_1i_2i_3} + \partial_i\log A\, \psi^{rs}_{i_1i_2i_3} - \frac{3i}{2}\,\imp\lbrace H_{i[i_1i_2}\,\tilde{\kappa}^{rs}_{i_3]} \rbrace +\frac{3i}{8}\,\imp\lbrace {}^{\star}H_{i[i_1i_2}\,\tilde{\kappa}^{rs}_{i_3]} \rbrace  \cr
&&- \frac{3}{8}\rep\lbrace {}^{\star}H_{i[i_1}{}^j\tilde{\psi}^{rs}_{i_2i_3]j} \rbrace + \frac{9i}{8}\,\imp\lbrace {}^{\star}H_{i[i_1i_2}\,\tkkappa^{rs}_{i_3]} \rbrace +\frac{9}{8}\,\rep\lbrace H_{i[i_1}{}^j\,\tilde{\psi}^{rs}_{i_2i_3]j} \rbrace \cr
&&= i\,{}^{\star}Y_{ii_1i_2i_3}{}^j\,\kappa^{rs}_j + 6i\,\delta_{i[i_1}\,Y_{i_2}\,\kkappa^{rs}_{i_3]} - \frac{3i}{8}\,\delta_{i[i_1}\,\imp\lbrace H_{i_2i_3]}{}^j \tilde{\kappa}^{rs}_j\rbrace +\frac{i}{2}\,\imp\lbrace H_{[i_1i_2i_3}\tilde{\kappa}^{rs}_{i]} \rbrace \cr
&& + \frac{9i}{8}\,\delta_{i[i_1}\,\imp\lbrace {}^{\star}H_{i_2i_3]}{}^j \tkkappa^{rs}_j \rbrace -\frac{3i}{2}\,\imp\lbrace {}^{\star}H_{[i_1i_2i_3}\tkkappa^{rs}_{i]} \rbrace~,
\label{ads4tcfh13}
\eea
\bea
{\cal D}^{\cal F}_i\tilde{\kappa}^{rs}_k &\defeq& \nabla_i\,\tilde{\kappa}^{rs}_k + (\partial_i\log A + iQ_i)\tilde{\kappa}^{rs}_i + 2i\,Y_i\,\tkkappa^{rs}_k + \frac{i}{4}\,\bar{H}_i{}^{j_1j_2}\,\psi^{[rs]}_{kj_1j_2} \cr
&&=i\,\delta_{ik}\,Y^j\,\tkkappa^{rs}_j + 2i\,Y_{[i}\,\tkkappa^{rs}_{k]} +\frac{i}{48}\,\delta_{ik}\,\bar{H}^{j_1j_2j_3}\,\psi^{[rs]}_{j_1j_2j_3} \cr
&& + \frac{i}{8}\,\bar{H}^{j_1j_2}{}_{[i}\,\psi^{[rs]}_{k]j_1j_2} + \frac{1}{8}\,{}^{\star}\bar{H}_{ik}{}^j\,\kkappa^{[rs]}_j - \frac{3}{8}\,\bar{H}_{ik}{}^j\,\kappa^{[rs]}_j~,
\label{ads4tcfh11}
\eea
\bea
{\cal D}^{\cal F}_i\tkkappa^{rs}_k &\defeq& \nabla_i\,\tkkappa^{rs}_k + (\partial_i\log A + iQ_i)\tkkappa^{rs}_i -2i\,Y_i\tilde{\kappa}^{rs}_k + \frac{i}{4}\,{}^{\star}\bar{H}_i{}^{j_1j_2}\,\psi^{[rs]}_{kj_1j_2} \cr
&&=-i\,\delta_{ik}\,Y^j\,\tilde{\kappa}^{rs}_j - 2i\,Y_{[i}\,\tilde{\kappa}^{rs}_{k]} +\frac{3i}{48}\,{}^{\star}\bar{H}^{j_1j_2j_3}\,\psi^{[rs]}_{j_1j_2j_3} \cr
&& +\frac{3i}{8}\,{}^{\star}\bar{H}^{j_1j_2}{}_{[i}\,\psi^{[rs]}_{k]j_1j_2} - \frac{1}{8}\,{}^{\star}\bar{H}_{ik}{}^j\,\kappa^{[rs]}_j - \frac{3}{8}\,\bar{H}_{ik}{}^j\,\kkappa^{[rs]}_j~,
\label{ads4tcfh12}
\eea
\bea
{\cal D}_i^{\cal F}\tilde{\psi}^{rs}_{i_1i_2i_3} &\defeq& \nabla_i\,\tilde{\psi}^{rs}_{i_1i_2i_3} + (\partial_i\log A + Q_i)\,\tilde{\psi}^{rs}_{i_1i_2i_3} + 3i\,{}^{\star}Y^{j_1j_2}{}_{i[i_1i_2}\,\tilde{\psi}^{rs}_{i_3]j_1j_2} \cr
&&-\frac{3i}{2}\, \bar{H}_{i[i_1i_2}\,\kappa^{(rs)}_{i_3]} + \frac{3i}{8}\, {}^{\star}\bar{H}_{i[i_1i_2}\kkappa^{(rs)}_{i_3]} - \frac{9}{8}\, \bar{H}^j{}_{i[i_1}\psi^{(rs)}_{i_2i_3]j}  +\frac{9i}{8}\,{}^{\star}\bar{H}_{i[i_1i_2}\,\kappa^{(rs)}_{i_3]} \cr
&&=\frac{i}{2}\,\delta_{i[i_1}\,{}^{\star}Y^{j_1j_2j_3}{}_{i_2i_3]}\,\tilde{\psi}^{rs}_{j_1j_2j_3} + 2i\,{}^{\star}Y^{j_1j_2}{}_{[ii_1i_2}\,\tilde{\psi}^{rs}_{i_3]j_1j_2} - \frac{3i}{8}\,\delta_{i[i_1}\, \bar{H}_{i_2i_3]}{}^j\, \kappa^{(rs)}_j  \cr
&&+\frac{i}{2}\,\bar{H}_{[i_1i_2i_3}\,\kappa^{(rs)}_{i]}  + \frac{9}{4}\,\delta_{i[i_2}\, \bar{H}^{j_1j_2}{}_{i_3}\,\psi^{(rs)}_{i_1]j_1j_2}  + \frac{9}{2}\, \bar{H}^j{}_{[i_1i_2}\,\psi^{(rs)}_{i_3i]j} \cr
&&+\frac{9i}{8}\,\delta_{i[i_1}\,{}^{\star}\bar{H}^j{}_{i_2i_3]}\,\kappa^{(rs)}_j - \frac{3i}{2}\, {}^{\star}\bar{H}_{[i_1i_2i_3}\,\kappa^{(rs)}_{i]}~.
\label{ads4tcfh14}
\eea
The (reduced) holonomy of the minimal connection ${\cal D}^{\cal F}$ is included in (the connected component of) $GL(72)\times GL(44)$.

\section{The TCFHs of warped AdS$_k$, $k\geq 5$, backgrounds}

\subsection{The TCFH of warped  AdS$_5$ backgrounds }

The fields of  warped AdS$_5$ backgrounds, AdS$_5\times N^5$, are
\begin{gather}
g = 2\,\ee^+\ee^- + (\ee^z)^2 +\sum_{a=1}^2 (\ee^a)^2   + g(N^5)~, \nn
F = Y\left [\ee^+\wedge \ee^-\wedge  \ee^z \wedge \ee^1 \wedge \ee^2 - \mathrm{dvol}(N^5) \right]~, \quad G = H
~,
\label{metric_ads5}
\end{gather}
where $Y$ is a function on $N^5$ and $H$ is a $U(1)$-twisted 3-form on $N^6$. The components $(\ee^+, \ee^-, \ee^z, \ee^i)$ of pseudo-orthonormal frame are defined as in the previous cases with the understanding that the warped factor $A$ is a function of $N^5$ and $\ee^i=e^i_I dy^I$ is an orthonormal frame on $N^5$, $g(N^5)=\delta_{ij} \ee^i \ee^j$, where $y^I$ are coordinates on $N^5$. Furthermore,  $\ee^a= A e^{z\over\ell} dx^a$, where $(u,r, z, x^a)$, $a=1,2$,  are the remaining coordinates of spacetime. The spacetime metric can be put into the standard warped form after a coordinate transformation.

As in previous cases, the KSEs of the theory can be integrated over the $(u, r, z, x^a)$ coordinates \cite{sbjggp} and the Killing spinors, $\epsilon$, can be expressed as,
$\epsilon=\epsilon(u,r,z, x^a, \sigma_\pm, \tau_\pm)$, where $\sigma_\pm$ and $\tau_\pm$ depend only on the coordinates of $N^5$ and $\Gamma_\pm\sigma_\pm=\Gamma_\pm\tau_\pm=0$. Again the integration over the $z$ coordinate introduces a new algebraic KSE on $\sigma_\pm$ and $\tau_\pm$ in addition to those
induced by the gravitino and dilatino KSEs of the theory. In particular, one finds that the gravitino KSE implies that $\nabla^{(\pm)}_i\sigma_{\pm} = 0$ and $\nabla^{(\pm)}_i\tau_{\pm} = 0$ along $N^5$, where the supercovariant connections are
\begin{align}\label{KSE5}
\nabla_{i}^{(\pm)} \equiv \nabla_i &\pm \frac{1}{2}\,\partial_i \log A  -\frac{i}{2}\,Q_i \pm \frac{i}{2}\, \Gamma_iY\Gamma_{x^1x^2z}+\left( -\frac{1}{96}\,(\Gamma\slashed{H})_i +\frac{3}{32}\,\slashed{H}_i \right)C*~,
\end{align}
and  the gamma matrices $\Gamma_{x^a}$, $a=1,2$, are considered in the frame $\ee^a$.

An argument similar to that used in the AdS$_3$ and AdS$_4$  cases leads to the conclusion that it suffices to consider the TCFH of only the $\sigma_+$ form bilinears. It is also known that if $\sigma_+$ is a Killing spinor, then $\Gamma_{x^1x^2}\sigma_+$ is also a $\sigma_+$-type of Killing spinor.  Moreover, if again $\sigma_+$ is a Killing spinor, then $ v^a \Gamma_{x^a} \Gamma_z \sigma_+$ is a $\tau_+$-type of Killing spinor for any constant vector $v$. After consideration of these properties of Killing spinors, one can conclude that it suffices  to consider the TCFH of the following  basis in the space of the form bilinears
\begin{gather}
\rho^{rs}=\langle \sigma^r, \sigma^s \rangle~, \quad \tilde{\rho}^{rs}=\langle \sigma^r, C\bar{\sigma}^s \rangle~, \nn
\kappa^{rs} = \langle \sigma^r, \Gamma_{x^1x^2z}\Gamma_i\,\sigma^s \rangle \,\ee^i~, \quad \tilde{\kappa}^{rs} = \langle \sigma^r, \Gamma_{x^1x^2z}\Gamma_i\,C\bar{\sigma}^s \rangle\, \ee^i~, \nn
\omega^{rs} = \frac{1}{2}\, \langle \sigma^r, \Gamma_{i_1i_2}\,\sigma^s \rangle\, \ee^{i_1}\wedge \ee^{i_2}~, \quad \tilde{\omega}^{rs} = \frac{1}{2}\, \langle \sigma^r, \Gamma_{i_1i_2}\,C\bar{\sigma}^s \rangle\, \ee^{i_1}\wedge \ee^{i_2} ~,
\label{ads5_bilinears}
\end{gather}
 where $\tilde{\rho}^{rs}, \tilde{\kappa}^{rs}, \rep \rho^{rs}, \rep \kappa^{rs}$ and $\imp\omega^{rs}$ are  symmetric while $\tilde{\omega}^{rs}, \imp \rho^{rs}, \imp \kappa^{rs}$ and $\rep\omega^{rs}$ are skew-symmetric in the exchange of $\sigma^r$ and $\sigma^s$ spinors and $\sigma_+=\sigma$. For example, the TCFH of the form bilinears that include $\langle \sigma^r, \Gamma_{z}\Gamma_i\,\sigma^s \rangle \,\ee^i$ and $v^a\langle \sigma^r, \Gamma_{a}\Gamma_i\,\sigma^s \rangle \,\ee^i$ can be easily computed form that of (\ref{ads5_bilinears}) form bilinears using the properties of the Killing spinors mentioned above.

 After a direct computation, the TCFH is
\bea
{\cal D}^{\cal F}_i \rho^{rs} &\defeq& \nabla_i\, \rho^{rs} = -\partial_i \log A \, \rho^{rs} + \frac{1}{8}\, \rep \{ {}^{\star}H_i{}^j\, \tilde{\kappa}^{rs}_j \} -\frac{3i}{16}\imp\{H_i{}^{j_1j_2}\,\tilde{\omega}^{rs}_{j_1j_2}\} ~,
\label{ads5tcfh1}
\eea
\bea
{\cal D}^{\cal F}_i \kappa_k^{rs} &\defeq& \nabla_i \,\kappa_k^{rs} + \partial_i \log A \, \kappa_k^{rs} -\frac{i}{2}\,\imp\{{}^{\star}H_i{}^j\,\tilde{\omega}_{kj}^{rs}\} \cr
&&= -i\, Y\, \omega_{ik}^{rs} + \frac{1}{8}\, \rep \{ {}^{\star}H_{ik}\, \tilde{\rho}^{rs} \} -\frac{3i}{16}\,\delta_{ik}\,\imp\{{}^{\star}H^{j_1j_2}\,\tilde{\omega}_{j_1j_2}^{rs}\} +\frac{3i}{4}\,\imp\{{}^{\star}H^j{}_{[i}\,\tilde{\omega}_{k]j}^{rs}\} \cr
&&-\frac{3}{8}\, \rep \{ H_{ik}{}^j\, \tilde{\kappa}_j^{rs} \}~,
\label{ads5tcfh2}
\eea
\bea
{\cal D}^{\cal F}_i \omega_{i_1i_2}^{rs} &\defeq& \nabla_i\, \omega_{i_1i_2}^{rs} + \partial_i \log A \, \omega_{i_1i_2}^{rs} -\rep\{ H^j{}_{i[i_1}\,\tilde{\omega}_{i_2]j}^{rs}\} + i\,\imp \{ {}^{\star}H_{i[i_1}\,\tilde{\kappa}_{i_2]}^{rs} \} \cr
&&= -2i\, Y\, \delta_{i[i_1}\kappa_{i_2]}^{rs} -\frac{1}{8}\,\delta_{i[i_1}\, \rep\{H_{i_2]}{}^{j_1j_2}\,\tilde{\omega}_{j_1j_2}^{rs}\} -\frac{3}{8}\,\rep\{H^{j}{}_{[i_1i_2}\,\tilde{\omega}_{i]j}^{rs}\}  \cr
&&+ \frac{3i}{4}\, \delta_{i[i_1}\,\imp\{ {}^{\star}H_{i_2]}{}^j\,\tilde{\kappa}_j^{rs} \} + \frac{9i}{8}\, \imp \{ {}^{\star}H_{[ii_1}\,\tilde{\kappa}_{i_2]}^{rs} \} + \frac{3i}{8}\,\imp \{ H_{ii_1i_2}\,\tilde{\rho}^{rs} \}~,
\label{ads5tcfh3}
\eea
\bea
{\cal D}^{\cal F}_i \tilde{\rho}^{rs} &\defeq& \nabla_i \tilde{\rho}^{rs} + (iQ_i + \partial_i \log A)\, \tilde{\rho}^{rs} =- i\, Y \tilde{\kappa}_i^{rs} + \frac{1}{8}\, {}^{\star}\HH_{i}{}^j\, \kappa_j^{(rs)} \cr
&&-\frac{3}{16}\, \HH_i{}^{j_1j_2}\,\omega^{(rs)}_{j_1j_2} ~,
\label{ads5tcfh4}
\eea
\bea
{\cal D}^{\cal F}_i \tilde{\kappa}_k^{rs} &\defeq& \nabla_i\, \tilde{\kappa}_k^{rs} + (iQ_i+\partial_i \log A)\, \tilde{\kappa}_k^{rs} - \frac{1}{2}\,{}^{\star}\HH_i{}^j\, \omega_{kj}^{(rs)} \cr
&&=-i\, Y\, \delta_{ik}\, \tilde{\rho}^{rs} + \frac{1}{8}\, {}^{\star}\HH_{ik}\, \rho^{(rs)}  -\frac{3}{16}\,\delta_{ik}\,{}^{\star}\HH{}^{j_1j_2}\,\omega_{j_1j_2}^{(rs)} + \frac{3}{4}\, {}^{\star}\HH{}^j{}{}_{[i}\,\omega_{k]j}^{(rs)}  \cr
&&- \frac{3}{8}\,\HH_{ik}{}^j  \kappa_j^{(rs)} ~,
\label{ads5tcfh5}
\eea
\bea
{\cal D}^{\cal F}_i\tilde{\omega}_{i_1i_2}^{rs} &\defeq& \nabla_i\,\tilde{\omega}_{i_1i_2}^{rs} + (iQ_i + \partial_i\log A)\,\tilde{\omega}_{i_1i_2}^{rs} - \HH{}^j{}_{i[i_1}\,\omega_{i_2]j}^{[rs]} + {}^{\star}\HH_{i[i_1}\kappa_{i_2]}^{[rs]} \cr
&&=-\frac{i}{2}\,{}^{\star}Y_{ii_1i_2}{}^{j_1j_2}\,\tilde{\omega}_{j_1j_2}^{rs} -\frac{1}{8}\,\delta_{i[i_1}\,\HH_{i_2]}{}^{j_1j_2}\,\omega_{j_1j_2}^{[rs]} -\frac{3}{8}\,\HH^j{}_{[i_1i_2}\omega_{i]j}^{[rs]} \cr
&& +\frac{3}{4}\,\delta_{i[i_1}\,{}^{\star}\HH_{i_2]}{}^j \kappa_j^{[rs]} + \frac{9}{8}\,{}^{\star}\HH_{[i_1i_2} \,\kappa_{i]}^{[rs]} + \frac{3}{8}\,\HH_{ii_1i_2} \,\rho^{[rs]} ~,
\label{ads5tcfh6}
\eea
where we have used that $(\prod_i \Gamma_i) \Gamma_{x^1 x^2z}\sigma_\pm=\pm \sigma_\pm$.
One can easily verify that the (reduced) holonomy of the minimal TCFH connection ${\cal D}^{\cal F}$  is included in (the connected component of) $U(1)\times SO(5)\times GL(35)\times GL(20)$.

\subsection{The TCFH of warped AdS$_6$ backgrounds}

For warped AdS$_6$ backgrounds, AdS$_6\times N^4$, the 5-form field strength $F$ vanishes, $F=0$, and the remaining fields are given as in (\ref{metric_ads5}), where now $a=1,2,3$. The pseudo-orthonormal frame is again given as in the AdS$_5$ case with the difference that there is an additional $\ee^a= A e^{z/\ell} dx^a$ frame, $\ee^3$, associated with a new coordinate $x^3$, and $\ee^i$ is an orthonormal frame on $N^4$.

The KSEs can again be integrated \cite{sbjggp} over the coordinates $(u,r, z, x^a)$ and the Killing spinors, $\epsilon$, can be expressed in terms of the spinors $\sigma_\pm$ and $\tau_\pm$ which have similar properties to those of AdS$_5$ backgrounds. Moreover, $\sigma_\pm$ and $\tau_\pm$ satisfy two algebraic KSEs, one is as a result of the gaugino KSE and the other arises during the integration over the $z$ coordinate. Furthermore, the gravitino KSE implies that $\nabla_i^{(\pm)}\sigma_{\pm}=0$ and $\nabla_i^{(\pm)}\tau_{\pm}=0$ on $N^4$, where the  supercovariant derivatives are
\begin{align}\label{KSE6}
\nabla_{i}^{(\pm)} \equiv \nabla_i &\pm \frac{1}{2}\,\partial_i \log A  -\frac{i}{2}\,Q_i +\left( -\frac{1}{96}\,(\Gamma\slashed{H})_i +\frac{3}{32}\,\slashed{H}_i \right)C*~.
\end{align}
It turns out that if $\sigma_+$ is a Killing spinor, then $v^a u^b \Gamma_{x^a x^b}\sigma_+$ is also a $\sigma_+$-type of Killing spinor for any constant vectors $v$ and $u$.  Also, if $\sigma_+$ is a Killing spinor, then $v^a \Gamma_{x^a} \Gamma_z \sigma_+$ is a $\tau_+$-type of Killing spinor for any constant vector $v$.

The TCFH factorises on the subspaces of even- and odd-degree form bilinears on $N^4$.  Because of the relation between the Killing spinors mentioned above, it suffices to consider the basis
\begin{gather}
\rho^{rs}=\langle \sigma^r, \sigma^s \rangle~, \quad \tilde{\rho}^{rs}=\langle \sigma^r, C\bar{\sigma}^s \rangle~, \nn
\rrho^{rs}=\langle \sigma^r, \Gamma_{(4)}\, \sigma^s \rangle~, \quad \mathring{\tilde{\rho}}^{rs}=\langle \sigma^r, \Gamma_{(4)}\, C\bar{\sigma}^s \rangle~, \nn
\omega^{rs} = \frac{1}{2}\, \langle \sigma^r, \Gamma_{i_1i_2}\, \sigma^s \rangle\,\ee^{i_1}\wedge \ee^{i_2}~, \quad \tilde{\omega}^{rs} = \frac{1}{2}\, \langle \sigma^r, \Gamma_{i_1i_2}\, C\bar{\sigma}^s \rangle\,\ee^{i_1}\wedge \ee^{i_2}~,
\label{ads6_bilinears}
\end{gather}
with $\Gamma_{(4)}=\Gamma_z\prod_{a=1}^3\Gamma_{x^a}$, in the space of even-degree form bilinears. Note that $\tilde{\rho}^{rs}$, $\trrho^{rs}$, $\rep \rho^{rs}$, $\rep \rrho^{rs}$ and $\imp\omega^{rs}$ are symmetric, while $\tilde{\omega}^{rs}$, $\imp \rho^{rs}, \imp \rrho^{rs}$ and $\rep\omega^{rs}$ are skew-symmetric in the exchange of $\sigma^r$ and $\sigma^s$ spinors.   The TCFH of the rest of even-degree form bilinears, e.g. of the form bilinears $\langle \sigma^r, v^a u^b \Gamma_{ab} \sigma^s \rangle$ and others, can be derived from that of (\ref{ads6_bilinears}).

 A direct computation of the TCFH of (\ref{ads6_bilinears}) form bilinears reveals that

\bea
{\cal D}^{\cal F}_i \rho^{rs} &\defeq& \nabla_i\, \rho^{rs} = -\partial_i \log A\,\rho^{rs} -\frac{1}{8}\,\rep\{ {}^{\star}H_i\,\trrho^{rs} \} - \frac{3i}{16}\,\imp\{H_i{}^{j_1j_2}\,\tilde{\omega}_{j_1j_2}^{rs}\} ~,
\label{ads6tcfh1}
\eea
\bea
{\cal D}^{\cal F}_i \rrho^{rs} &\defeq& \nabla_i\, \rrho^{rs} = -\partial_i \log A\,\rrho^{rs} -\frac{1}{8}\,\rep\{ {}^{\star}H_i\,\tilde{\rho}^{rs} \} + \frac{3i}{8}\,\imp\{{}^{\star}H^j\,\tilde{\omega}_{ij}^{rs}\}~,
\label{ads6tcfh2}
\eea
\bea
{\cal D}^{\cal F}_i \omega_{i_1i_2}^{rs} &\defeq& \nabla_i\, \omega_{i_1i_2}^{rs} +\partial_i \log A\, \omega_{i_1i_2}^{rs} - \rep\{H^j{}_{i[i_1}\tilde{\omega}_{i_2]j}^{rs}\} \cr
&&= -\frac{3}{8}\,\rep\{H^j{}_{[i_1i_2}\,\tilde{\omega}^{rs}_{i]j}\} -\frac{1}{8}\,\delta_{i[i_1}\,\rep\{H_{i_2]}{}^{j_1j_2}\,\tilde{\omega}^{rs}_{j_1j_2}\} -\frac{3i}{4}\, \delta_{i[i_1}\, \imp\{ {}^{\star}H_{i_2]}\,\trrho^{rs} \}\cr
&& + \frac{3i}{8}\,\imp \{ H_{ii_1i_2}\,\tilde{\rho}^{rs} \}~,
\label{ads6tcfh3}
\eea
\bea
{\cal D}^{\cal F}_i \tilde{\rho}^{rs} &\defeq& \nabla_i\,\tilde{\rho}^{rs} +(iQ_i+\partial_i\log  A)\, \tilde{\rho}^{rs}= - \frac{1}{8}\,{}^{\star}\HH_i\,\rrho^{(rs)} - \frac{3}{16}\,\HH_i{}^{j_1j_2}\,\omega_{j_1j_2}^{(rs)}~,
\label{ads6tcfh4}
\eea
\bea
{\cal D}^{\cal F}_i \trrho^{rs} &\equiv \nabla_i\,\trrho^{rs} +(iQ_i+\partial_i\log  A)\, \trrho^{rs} =- \frac{1}{8}\,{}^{\star}\HH_i\, \rho^{(rs)} + \frac{3}{8}\,{}^{\star}\HH^j\,\omega_{ij}^{(rs)}~,
\label{ads6tcfh5}
\eea
\bea
{\cal D}^{\cal F}_i \tilde{\omega}_{i_1i_2}^{rs} &\defeq& \nabla_i\,\tilde{\omega}_{i_1i_2}^{rs} + (iQ_i+\partial_i\log A)\,\tilde{\omega}^{rs}_{i_1i_2} - \HH^j{}_{i[i_1}\,\omega_{i_2]j}^{[rs]} \cr
&&= -\frac{3}{8}\,\HH^j{}_{[i_1i_2}\,\omega_{i]j}^{[rs]} - \frac{1}{8}\,\delta_{i[i_1}\,\HH_{i_2]}{}^{j_1j_2}\,\omega^{[rs]}_{j_1j_2} -\frac{3}{4}\, \delta_{i[i_1}{}^{\star}\HH_{i_2]}\,\rrho^{[rs]} \cr
&&+ \frac{3}{8}\,\HH_{ii_1i_2}\,\rho^{[rs]}~,
\label{ads6tcfh6}
\eea
where we have used that $(\prod_i \Gamma_i) \Gamma_{x^1x^2 x^3z}\sigma_\pm=\pm \sigma_\pm$.
The (reduced) holonomy  of the minimal TCFH connection ${\cal D}^{\cal F}$ is included in (the connected component of) $U(1)\times SO(4)\times GL(18)$.

Next, a basis in the space of odd-degree form bilinears is
\bea
&&\kappa=\langle \sigma^r, \Gamma_{zi} \sigma^s \rangle\, \ee^i~,~~~\mathring{\kappa}=\langle \sigma^r, \Gamma_{x^1x^2x^3i} \sigma^s \rangle\, \ee^i~,
\cr
&&\tilde \kappa=\langle \sigma^r, \Gamma_{zi} C\bar{\sigma}^s \rangle\, \ee^i~,~~~\mathring{\tilde\kappa}=\langle \sigma^r, \Gamma_{x^1x^2x^3 i} C\bar{\sigma}^s \rangle\, \ee^i~,
\eea
where $\tkkappa$, $\imp\kappa$ and $\rep\kkappa$ are symmetric while $\tilde{\kappa}$, $\rep\kappa$ and $\imp\kkappa$ are skew-symmetric in the exchange of the spinors $\sigma^r$ and $\sigma^s$. There are more odd-degree form bilinears that one can consider but their TCFH can be computed from the one of the basis above. The TCFH reads
\bea
{\cal D}^{\cal F}\kappa^{rs}_k &\defeq& \nabla_i\,\kappa^{rs}_k + \partial_i\log A\,\kappa^{rs}_k - \frac{i}{2}\,\imp\lbrace {}^{\star}H_i\,\tkkappa^{rs}_k \rbrace \cr
&&= - \frac{3i}{8}\,\delta_{ik}\,\imp\lbrace {}^{\star}H^j\,\tkkappa^{rs}_j \rbrace - \frac{3i}{4}\,\imp\lbrace {}^{\star}H_{[i}\,\tkkappa^{rs}_{k]} \rbrace -\frac{3}{8}\,\rep\lbrace H_{ik}{}^j\,\tilde{\kappa}^{rs}_j \rbrace~,
\label{ads6tcfh7}
\eea
\bea
{\cal D}^{\cal F}\kkappa^{rs}_k &\defeq& \nabla_i\,\kkappa^{rs}_k + \partial_i\log A\,\kkappa^{rs}_k + \frac{i}{2}\,\imp\lbrace {}^{\star}H_i\,\tilde{\kappa}^{rs}_k \rbrace \cr
&&= \frac{3i}{8}\,\delta_{ik}\,\imp\lbrace {}^{\star}H^j\,\tilde{\kappa}^{rs}_j \rbrace + \frac{3i}{4}\,\imp\lbrace {}^{\star}H_{[i}\,\tilde{\kappa}^{rs}_{k]} \rbrace -\frac{3}{8}\,\rep\lbrace H_{ik}{}^j\,\tkkappa^{rs}_j \rbrace~,
\label{ads6tcfh8}
\eea
\bea
{\cal D}^{\cal F}_i\tilde{\kappa}^{rs}_k \defeq \nabla_i\,\tilde{\kappa}^{rs}_k + (\partial_i\log A + iQ_i)\,\tilde{\kappa}^{rs}_k = -\frac{3}{8}\,\bar{H}_{ik}{}^j\,\kappa^{[rs]}_j~,
\label{ads6tcfh9}
\eea
\bea
{\cal D}^{\cal F}_i\tkkappa^{rs}_k \defeq \nabla_i\,\tkkappa^{rs}_k + (\partial_i\log A + iQ_i)\,\tkkappa^{rs}_k = -\frac{3}{8}\,\bar{H}_{ik}{}^j\, \kkappa^{(rs)}_j~.
\label{ads6tcfh10}
\eea
The (reduced) holonomy of the minimal connection ${\cal D}^{\cal F}$ is included in  (the connected component of) $\times^2 GL(12)\times SO(4)$.

\section{TCFHs and hidden symmetries}

\subsection{Symmetries of a  spinning particle probe}

A consequence of the TCFH is that the form bilinears of supersymmetric backgrounds satisfy a generalisation of the CKY equation with respect to the TCFH connection \cite{tcfhgp}.
This indicates that the form bilinears may generate (hidden) symmetries for certain probes propagating on these backgrounds. This question has been investigated in \cite{ebgp1, ebgp2, lgjpgp, ebgp3}. Here we shall explore the question on whether the TCFH on the internal spaces of AdS backgrounds generate symmetries for spinning particle probes. This will be illustrated with examples that include the maximally supersymmetric AdS$_5$ solution as well as some other AdS$_2$ and AdS$_3$ solutions that arise as near horizon geometries of intersecting IIB branes, see \cite{ggpt, cham, kallosh, boonstra}.

As in all examples we shall consider the warp factor $A$ is constant, the dynamics of a spinning particle propagating on such an  AdS background factorises into one part that involves the dynamics of the probe on the AdS subspace and another part that involves the dynamics of the probe  on the internal space. Focusing on the latter, the action of such a spinning particle probe can be described by the action
\bea \label{probe_action}
A = -\frac{i}{2}\int d\tau\, d\theta \,g_{IJ}\, Dy^I\, \partial_{\tau}y^J~,
\eea
where $y=y(\tau,\theta)$ is a superfield with $\tau$ and $\theta$ the even and odd coordinates of the worldline superspace, and $D$ is the superspace derivative satisfying $D^2=i\partial_{\tau}$.

The symmetries of (\ref{probe_action}) that concern us here are those generated by forms on the internal space $N$. Given such a form $\beta$ the above action is invariant under the infinitesimal transformation
\bea \label{probe_transformation}
\delta y^I = \alpha\, \beta^I{}_{J_1\dots J_{k-1}}Dy^{J_1}\cdots Dy^{J_{k-1}}~,
\eea
 provided $\beta$ is a KY form, where $\alpha$ is an infinitesimal parameter.

 It is clear that not all Killing spinor form bilinears generate symmetries for the action (\ref{probe_action}). This is because although they are CKY forms with respect to the TCFH connection,  they are not KY forms which is more restrictive.  However, we shall demonstrate in many examples below that the TCFH simplifies on special supersymmetric backgrounds and the form bilinears become KY (or CCKY) forms which in turn generate symmetries for the  action (\ref{probe_action}).

\subsection{The maximally supersymmetric AdS$_5$  solution}

The only non-vanishing form field strength of the AdS$_5\times S^5$ maximally supersymmetric  solution is the 5-form flux $F$ which is determined in terms of the (constant) function $Y$ on the internal space $S^5$.  The IIB scalars as well as the warped factor $A$ are constant. Also,  without loss of generality, one can set $A=1$. In this case, the TCFH dramatically simplifies and  yields
\bea
&&{\cal D}^{\cal F}_i \rho^{rs} \defeq \nabla_i\, \rho^{rs} = 0 ~,~~~
{\cal D}^{\cal F}_i \kappa_k^{rs} \defeq \nabla_i \,\kappa_k^{rs} = -i\, Y\, \omega_{ik}^{rs}~,
\cr
&&{\cal D}^{\cal F}_i \omega_{i_1i_2}^{rs} \defeq \nabla_i\, \omega_{i_1i_2}^{rs} = -2i\, Y\, \delta_{i[i_1}\kappa_{i_2]}^{rs}~,~~~
{\cal D}^{\cal F}_i \tilde{\rho}^{rs} \defeq \nabla_i \tilde{\rho}^{rs}  =- i\, Y \tilde{\kappa}_i^{rs}~,
\cr
&&
{\cal D}^{\cal F}_i \tilde{\kappa}_k^{rs} \defeq \nabla_i\, \tilde{\kappa}_k^{rs} =-i\, Y\, \delta_{ik}\, \tilde{\rho}^{rs}~,~~~
{\cal D}^{\cal F}_i\tilde{\omega}_{i_1i_2}^{rs} \defeq \nabla_i\,\tilde{\omega}_{i_1i_2}^{rs} =-\frac{i}{2}\,{}^{\star}Y_{ii_1i_2}{}^{j_1j_2}\,\tilde{\omega}_{j_1j_2}^{rs}~.
\eea
Clearly, the (reduced) holonomy of the TCFH connection is included in $SO(5)$. Furthermore, $\kappa$, ${}^\star\omega$, ${}^\star \tilde \kappa$ and $\tilde \omega$ are KY forms on $S^5$ and so generate symmetries for the spinning particle action (\ref{probe_action}), where the Hodge duality operation has been taken over $S^5$. As the IIB scalars are constant, the $U(1)$ twist of $\tilde\rho$, $\tilde \kappa$ and $\tilde \omega$ vanishes and all of them are (standard) forms on $S^5$.

\subsection{AdS$_3$ solution from strings on 5-branes}

Taking the IIB 5-form flux to vanish and the IIB scalars to be constant, an ansatz that includes the near horizon geometry of a fundamental (D-) string on a NS5- (D5-) brane is \bea
g = g_{\ell}(AdS_3) + g(S^3) + g(\mathbb{R}^4)~, \quad G = p\,\mathrm{dvol}_{\ell}(AdS_3) + q\,\mathrm{dvol}(S^3)~,
\eea
where  $g_{\ell}(AdS_3)$ ($g(S^3)$) and $\mathrm{dvol}_{\ell}(AdS_3)$  ($\mathrm{dvol}(S^3)$) is the standard metric  and associated volume form on AdS$_3$  ($S^3$) with radius $\ell$ (unit radius), respectively,   $g(\mathbb{R}^4)$ is the Euclidean metric of $\mathbb{R}^4$ and $p,q \in \bC$. As the 5-form vanishes and the IIB scalars are constant, one has $Y=0$ and $\xi=Q=0$.  Moreover, without loss of generality, one can set $A=1$. From the ansatz above  $H=q\,\mathrm{dvol}(S^3)$ and $\Phi = q$. See \cite{ggpt, cham, kallosh, boonstra} for an extensive discussion of the near horizon geometries of intersecting branes \cite{gppkt1, jpg3, aat}.

To  determine the constants\footnote{We use the approach of \cite{sbjggp} to investigate the KSEs of AdS backgrounds as it has the advantage of deriving the results from first principles without any additional assumptions,  like for example the factorisation  the Killing spinors.} $p,q$ and $\ell$, the field equation\footnote{This corrects a sign in the field equation for $\xi$ in \cite{sbjggp} for warped AdS$_3$ backgrounds. Although a modification in the analysis of some cases in \cite{slgp} is needed, it does not not affect the final conclusion.}  of the IIB 1-form flux, $H^2 = 6\Phi^2$, gives  $q^2 = p^2$.  Next,  the Einstein field equation along $S^3$ and the warp factor field equation
\begin{gather}
R^{S^3}_{\alpha\beta} = \frac{1}{4}\,\bar{H}_{(\alpha}{}^{\gamma\zeta}H_{\beta)\gamma\zeta} + \frac{1}{8}\lVert\, \Phi\, \rVert^2 \delta_{\alpha\beta} - \frac{1}{48}\,\lVert\, H\, \rVert^2\delta_{\alpha\beta}~, \nn
\frac{3}{8}\, \lVert \, \Phi \, \rVert^2 + \frac{1}{48}\, \lVert\, H \, \rVert^2 - 2\ell^{-2} = 0~,
\end{gather}
respectively, give
 $\ell^2=1$ and $|p|^2=4$, i.e. the AdS$_3$ and $S^3$ subspaces have the same radius.

The  dilatino KSE, ${\cal A}^{(+)}\sigma_+=0$, with
\bea
{\cal A}^{(+)} = -\frac{1}{4}\,\Phi\Gamma_z + \frac{1}{24}\,\slashed{H}~,
\eea
  gives the condition $\Gamma_z\Gamma_{(3)}\sigma_+=(q/p)\sigma_+$, where $\Gamma_{(3)}$ is the product of the three gamma matrices along the orthonormal directions tangent to the 3-sphere.  The additional  algebraic KSE \cite{sbjggp}, $\Xi_+\sigma_+=0$, with
\bea
\Xi_+ = -\frac{1}{2\ell} + \left( \frac{1}{96}\,\Gamma_z\slashed{H} + \frac{3}{16}\,\Phi \right)C*~,
\eea
which arises from the integration of gravitino KSE along $z$ , yields the relation $C\bar{\sigma}_+ =  (2/q)\sigma_+$.  Therefore $|q|= 2$ as expected.

Furthermore, the the gravitino KSE along $\bR^4$ implies that the Killing spinors $\sigma_+$ do not depend on the coordinates of $\bR^4$. Using these, the gravitino KSE along $S^3$ can be written as
\bea
\nabla^{(+)}_\alpha = \nabla^{S^3}_\alpha - \frac{1}{2}\Gamma_z\Gamma_\alpha ~,
\eea
 and does not impose any additional conditions on $\sigma_+$, where we have used both $\Gamma_z\Gamma_{(3)}\sigma_+=(q/p)\sigma_+$ and $C\bar{\sigma}_+ =  (2/q)\sigma_+$. As a consequence, there are no additional conditions on $p$ and $q$ and therefore there is a solution for any $p\in \bC$ such that $|p|=2$ and $q=\pm p$. From the analysis above, it is clear that the KSEs on $\sigma_+$ admit 4 linearly independent solutions. This is also the case  for the KSEs on the remaining $\sigma_-$ and $\tau_\pm$ spinors. As a result, all these solutions admit 16 Killing spinors, i.e. they preserve $1/2$ of supersymmetry as expected.

 Next consider the form bilinears with components only along $S^3$. Because of $C\bar{\sigma}_+ =  (2/q)\sigma_+$, the $\tilde \phi$ bilinears are not linearly independent from the $\phi$ bilinears, where $\phi$ stands for all bilinears. It is easy to see that  $\kappa$ is a KY form, while $\psi$ and $\omega$ are CCKY forms. Therefore ${}^*\psi$ and ${}^*\omega$ are also KY forms, where the duality operation has been taken over $S^3$. Hence, $\kappa$ and  ${}^*\omega$ generate symmetries for the particle action\footnote{For the near horizon geometry of a fundamental string on a NS5-brane, one can consider other probes like a spinning particle probe with a 3-form coupling  as well as a fundamental string probe with a Wess-Zumino term. In such a case, the form bilinears are covariantly constant with respect to a connection with torsion and generate symmetries for these probe actions \cite{wphgp}.} (\ref{probe_action}) restricted on $S^3$.

\subsection{AdS$_3$ solution from two intersecting D3-branes}

An ansatz which includes the near horizon geometry of two D3-branes intersecting on a 1-brane is
\bea
g=g_\ell(AdS_3)+ g(\bR^4)+ g(S^3)~,~~~F = \mathrm{dvol}_{\ell}(AdS_3)\wedge Y - {}^{\star_7} Y~,
\eea
where $H, \Phi$  vanish, the scalar fields are constant and so $Q,\xi=0$, $Y=p \,dx^1\wedge dx^2 + q\,dx^3\wedge dx^4$, $p,q\in \bR$, is a 2-form on $\mathbb{R}^4$ with Cartesian coordinates $(x^1,\dots ,x^4)$. The metrics $g_\ell(AdS_3)$, $g(\bR^4)$ and $g(S^3)$, and volume form $\mathrm{dvol}_{\ell}(AdS_3)$  have already been described in the previous example. We have also set $A=1$. To specify the solution, we have to determine the parameters $\ell, p$ and $q$ of the ansatz.

 The field equation of the warp factor, $Y^2=\ell^{-2}$, as well as the Einstein field equation, $R^{(7)}_{ij}= 2 Y^2 \delta_{ij}-8 Y^2_{ij}$, restricted along $\bR^4$ give $p^2+q^2=1/2$ and $\ell=1$, i.e.   AdS$_3$ has the same radius as $S^3$. The algebraic KSE \cite{sbjggp}, $\Xi^{(+)}\sigma_+=0$,  has solutions provided that $\Gamma_{12} \sigma_+=-i \lambda \sigma_+$, $\Gamma_{34} \sigma_+=-i \mu \sigma_+$  and that $\lambda p+\mu q=1$, where $\lambda, \mu=\pm 1$. Using this equation together with the gravitino KSE along $\bR^4$, one finds that
 $p=\lambda/2$ and $q=\mu/2$.

Furthermore, the supercovariant derivative along $S^3$ is
\bea
\nabla^{(+)}_\alpha = \nabla^{S^3}_\alpha -{1\over2} \Gamma_z\Gamma_\alpha~,
\label{xxxyy}
\eea
and the associated KSE does not impose any additional conditions on $\sigma_+$.  As a consequence, the KSEs on $\sigma_+$  admit  4 linearly independent solutions.  A similar analysis reveals that this is the case for the remaining KSEs on $\sigma_-$ and $\tau_\pm$.   Thus the background preserves 16 supersymmetries.

Considering the form bilinears along $S^3$,
a direct computation of the TCFH connection using (\ref{xxxyy})  reveals that  $\kappa$ and $\tilde{\kappa}$ are KY forms, $\omega$ and $\tilde{\omega}$ are CCKY forms, and $\psi$ and $\tilde{\psi}$ are parallel, i.e. the latter are proportional to the volume form of $S^3$. As a consequence, all of them or their duals on $S^3$ generate symmetries for the probe action (\ref{probe_action}).

\subsection{AdS$_2$ solution from four intersecting D3-branes}

An ansatz that includes the near horizon geometry of four intersecting D3-branes on a 0-brane solution  is
\bea
g = g_{\ell}(AdS_2) + g(S^2) + g(\mathbb{R}^6)~, \quad F = \mathrm{dvol}_{\ell}(AdS_2) \wedge Y + {}^{\star_8}Y~,
\eea
with $H, \Phi, \xi, Q=0$, i.e.  the scalar fields are constant,
where
\bea
&&Y=p\, dx^1 \wedge dx^2 \wedge dx^3 + q\, dx^1 \wedge dx^4 \wedge dx^5 + r\, dx^2 \wedge  dx^4 \wedge dx^6
\cr
&&\qquad\qquad+ s\, dx^3 \wedge  dx^5 \wedge dx^6~,
 \eea
 $p,q,r,s\in \bR$,  is a 3-form on $\mathbb{R}^6$ with  Cartesian coordinates $(x^1, \dots, x^6)$.  The metrics $g_{\ell}(AdS_2)$, $g(S^2)$ and $g(\mathbb{R}^6)$ and volume form $\mathrm{dvol}_{\ell}(AdS_2)$ are defined in an analogous way to those described for the  AdS$_3$ backgrounds in previous sections. Again, we set $A=1$.

To find the values of the constants $p,q,r,s,\ell$ such that the above ansatz is a solution, consider the Einstein equation
$R^{(8)}_{ij}=- 4 Y^2_{ij}+2/3\, \delta_{ij} Y^2$.  In particular restricting this equation on $\bR^6$, we find that $p^2=q^2=r^2=s^2$.  Furthermore, the warp factor field equation $2/3 Y^2=\ell^{-2}$ gives $16 p^2=\ell^{-2}$.  Next restricting the Einstein equation on $S^2$, we have that $\ell=1$ which in turn gives
$p^2=q^2=r^2=s^2=1/16$. This specifies the solution.

It remains to count the number of supersymmetries preserved by the background. Restricting the gravitino KSE
\bea
\nabla_i^{(+)}\eta_+=\nabla_i \eta_+-{i\over4} \slashed{Y}_i\eta_++ {i\over12} \slashed {\Gamma Y}_i\eta_+=0
\eea
 along $\bR^6$, we get the conditions
 \bea
\left( p\Gamma_{23} + q\Gamma_{45} - r\Gamma_{1246} -s\Gamma_{1356} \right)\eta_+ &=0~, \cr
\left( p\Gamma_{31} + r\Gamma_{46} - q\Gamma_{2145} - s\Gamma_{2356} \right)\eta_+ &=0~, \cr
\left( p\Gamma_{12} +s\Gamma_{56} -q\Gamma_{3145} -r\Gamma_{3246} \right)\eta_+ &=0~.
\label{ads2s2r6conditions}
\eea
These can be solved by decomposing $\eta_+$ into the eigenspaces of $\Gamma_{2345}$ and $\Gamma_{1346}$ as  $\Gamma_{2345}\eta_+=\lambda\eta_+$, and $\Gamma_{1346}\eta_+=\zeta\eta_+$, where $\lambda,\zeta=\pm 1$.  In such a case, the above equations can be solved to find
\bea
q=-\lambda p~,~~~r=\zeta p~,~~~s=\zeta \lambda p~.
\eea
Clearly, there are solutions to the field equations which are not supersymmetric.
Next, the gravitino KSE along $S^2$ yields
\bea
\nabla_\alpha^{S^2} \eta_++2i p \Gamma_\alpha \Gamma_{123} \eta_+=0~,
\eea
and does not impose any additional conditions on $\eta_+$.  Therefore, the KSEs on $\eta_+$ have 4 linearly independent solutions. A similar analysis reveals that the KSEs on $\eta_-$ have also 4 linearly independent solutions. As a result, the background preserves 1/4 of supersymmetry as expected.

Considering the form bilinears restricted on $S^2$, it is easy to see that $\omega$ is a KY form while $\tilde{\omega}$ is a parallel form on $S^2$ and so the latter is proportional to the volume form. Both generate symmetries for the spinning particle action (\ref{probe_action}).

\section{Concluding Remarks}

We have presented the TCFHs on the internal space of all IIB  AdS backgrounds. Therefore, we have demonstrated that all Killing spinor form bilinears satisfy the CKY equation with respect to the TCFH connection.  We have also investigated some of the properties of the TCFHs we have found, like for example the (reduced) holonomy of the TCFH connections. Moreover, we have  given  some examples of solutions for which  the form bilinears are KY and CCKY forms and therefore generate symmetries for spinning particle probes propagating on the internal spaces of these backgrounds.  These solutions  include the maximally supersymmetric AdS$_5$ solution as well as the near horizon geometries of some intersecting IIB branes.

Although we have presented some key examples which illustrate the close relationship between TCFHs  and  symmetries for certain particle probes propagating on supersymmetric backgrounds, this investigation has proceeded on a case by case basis.  In particular, there is not a systematic way to relate the conditions on the Killing spinor form bilinears described by the TCFH with the invariance conditions of certain probes propagating on the associated supersymmetric backgrounds.
Although the TCFHs are  determined by the KSEs of the supergravity theory under investigation given a choice of form bilinears and that of the TCFH connection, there is a plethora of actions with different couplings and worldline fields that describe the dynamics of spinning particle type of probes propagating on supersymmetric backgrounds, see \cite{rcgp}.  Each such action gives rise to different invariance conditions for transformations generated by Killing spinor form bilinears. Although some such probe actions have been considered before in this context \cite{ebgp1, lgjpgp}, a systematic understanding of the relation between TCFHs and invariance conditions for probe actions is still missing, and it will be considered  in the future.


\newpage

\setcounter{section}{0}
\setcounter{subsection}{0}
\setcounter{equation}{0}

\begin{appendices}

\section{Notation and conventions}

Let $\phi$ be a $k$-form  $\phi\in\Omega^k(M)$ on a n-dimensional manifold $N$ with metric $g$. Then
\bea
\phi = \frac{1}{k!}\,\phi_{i_1\dots i_k}\,\ee^{i_1}\wedge \cdots \wedge \ee^{i_k}~,
\eea
and the components of its exterior derivative, $d\phi$,  are
$
(d\phi)_{i_1\dots i_{k+1}} = (k+1)\,\nabla_{[i_1}\phi_{i_2\dots i_{k+1}]}$,
where $i=1,\dots,n$. The components of the  Hodge dual, ${}^\star\phi$, of $\phi$ are
\bea
{}^{\star}\phi_{i_1\dots i_{n-k}} = \frac{1}{k!}\,\phi_{j_1\dots j_k}\epsilon^{j_1\dots j_k}{}_{i_1\dots i_{n-k}}~,
\eea
where $\epsilon$ is the Levi-Civita tensor.
Note that $\phi$ is self-dual if ${}^{\star}\phi=\phi$, and anti-self-dual if ${}^{\star}\phi=-\phi$. Furthermore, for $\phi$ complex, we have, $\lVert\, \phi\, \rVert^2 = \bar{\phi}_{i_1\dots i_k}\phi^{i_1\dots i_k}$, 
and $\phi^2=\phi_{i_1\dots i_k} \phi^{i_1\dots i_k}$.

The Clifford algebra element associated with a form $\phi$ is
\bea
\fsl{\phi} = \phi_{i_1\dots i_k}\Gamma^{i_1\dots i_k}~,
\eea
and
\bea
\fsl{\phi}_{i_1} = \phi_{i_1 i_2\dots i_k}\Gamma^{i_2\dots i_k}~, \quad (\Gamma\fsl{\phi})_{i_1} = \Gamma_{i_1}{}^{i_2\dots i_{k+1}}\phi_{i_2\dots i_{k+1}}~,
\eea
where $\Gamma_i$ is a basis in the Clifford algebra, $\Gamma_i \Gamma_j+\Gamma_j \Gamma_i=2 \delta_{ij} {\bf 1}$.

\section{Complete integrability of AdS geodesic flow}

It is well known that the geodesic flow equations on AdS$_n$ are separable and can be integrated. Here we shall prove the Liouville integrability of the geodesic flow by explicitly presenting the independent charges in involution.    It is well-known that  AdS$_n$, $n\geq2$, can be described as  hyper-surface
\bea
\eta_{ab} x^a x^b=-\ell^2~,
\eea
in $\bR^{n-1,2}$, where $\eta$ is the mostly plus signature standard metric on  $\bR^{n-1,2}$ and $\ell$ is the radius. The metric on AdS$_n$ is the restriction of $\eta$ on the hyper-surface.    The  Killing vector fields on AdS$_n$ written in $\bR^{n-1,2}$ Cartesian coordinates are
\bea
k_{ab}=x_a \partial_b-x_b\partial_a~,
\label{adsrot}
\eea
where $x_a=\eta_{ab} x^b$. Observe that $k_{ab}$ are orthogonal to the radial direction $x^c$. Setting $Q_{ab}=x_a p_b-x_b p_a$, the $n$  conserved charges
\bea
D_m={1\over4} \sum_{a,b \geq n+2-m}  (Q_{ab})^2~,~~~m=2,\dots, n+1~.
\label{chads}
\eea
are independent and in involution.  Therefore, the geodesic flow on AdS$_n$ is completely integrable as expected.
Observe that $-D_{n+1}$ is the Hamiltonian of the geodesic system on AdS$_n$ as
\bea
-D_{n+1}=-{1\over4} (x_a p_b-x_b p_a) (x^a p^b- x^b p^a)=-{1\over2} \eta_{ab} x^a x^b \eta^{cd} p_c p_d={\ell^2\over2}  \eta^{cd} p_c p_d~,
\eea
where we have used that $x^a p_a=0$.

As the geodesic equation on AdS$_k \times S^m\times \bR^n$ factorises into those on AdS$_k$, $S^m$ and $\bR^n$, respectively, the Liouville integrability of the geodesic flow on AdS$_k \times S^m\times \bR^n$  reduces to that of the geodesic flow on each of the three subspaces.
The Liouville integrability of the geodesic flow on AdS$_k$ has been demonstrated above and that of the round $S^m$ has been considered before; for the conserved charges in involution see \cite{ebgp2, lgjpgp}. This demonstrates that the geodesic flow on all AdS$_k \times S^m\times \bR^n$ backgrounds is Liouville integrable.

\section{The TCFH of IIB theory}

 In \cite{lgjpgp}, we have given the TCFH of IIB supergravity in the string frame. As  we have used the Einstein frame for determining the TCHFs of IIB AdS backgrounds,  we also present the TCFH of IIB theory in Einstein frame for completeness. A basis in the space of form bilinears, up to a Hodge duality, can be chosen as
\begin{gather}
k^{rs} = \langle \epsilon^r, \Gamma_P \, \epsilon^s \rangle_D \, e^P~,  \quad
\tilde{k}^{rs} = \langle \epsilon^r, \Gamma_P \, C\bar{\epsilon}^s \rangle_D \, e^P~, \nn
\pi^{rs} = \frac{1}{3!} \, \langle \epsilon^r, \Gamma_{P_1 P_2 P_3} \, \epsilon^s \rangle_D \, e^{P_1}\wedge e^{P_2} \wedge e^{P_3} ~, \quad \tilde{\pi}^{rs} = \frac{1}{3!} \, \langle \epsilon^r, \Gamma_{P_1 P_2 P_3} \, C\bar{\epsilon}^s \rangle_D \, e^{P_1}\wedge e^{P_2} \wedge e^{P_3} ~, \nn
\tau^{rs} = \frac{1}{5!} \, \langle \epsilon^r, \Gamma_{P_1\dots P_5} \, \epsilon^s \rangle_D \, e^{P_1}\wedge\dots\wedge e^{P_5}, \quad \tilde{\tau}^{rs} = \frac{1}{5!} \, \langle \epsilon^r, \Gamma_{P_1\dots P_5} \, C\bar{\epsilon}^s \rangle_D \, e^{P_1}\wedge\dots\wedge e^{P_5} ~,
\label{iib_bilinears}
\end{gather}
where $\langle \cdot, \cdot \rangle_D$ is the Dirac inner product, $e^P$ is a spacetime frame and $\epsilon^r$ is a $\mathfrak{spin}(9,1)$ complex Weyl spinor, obeying the chirality condition $\Gamma_{0\dots 9}\,\epsilon^r=\epsilon^r$. The gravitino KSE of IIB supergravity, ${\cal D}_M \epsilon^r=0$, is the parallel transport equation of the supercovariant derivative
\bea\label{KSE}
{\cal D}_M \equiv \tilde{\nabla}_M + \frac{i}{48}\Gamma^{N_1\dots N_4} F_{N_1\dots N_4M} - \frac{1}{96}\left(\Gamma_M{}^{N_1N_2 N_3}G_{N_1 N_2 N_3}-9\Gamma^{N_1N_2}G_{MN_1N_2}\right)C*~,
\eea
where
\begin{equation}
\tilde{\nabla}_M = D_M + \frac{1}{4}\Omega_{M,AB}\Gamma^{AB}~, \quad D_M = \partial_M - \frac{i}{2}Q_M~,
\end{equation}
is the spin connection, $\nabla_M = \partial_M + \frac{1}{4}\Omega_{M,AB}\Gamma^{AB}$, twisted with a real $U(1)$ connection $Q$ that depends on the IIB scalars. Moreover, $F$ is real, whereas $G$ is complex. We choose the spacetime orientation as $\epsilon_{0\dots 9}=1$ and the self-duality condition on $F$ is expressed as $F_{M_1\dots M_5}=-\frac{1}{5!}\epsilon_{M_1\dots M_5}{}{}{}{}{}^{N_1\dots N_5}F_{N_1\dots N_5}$. The TCFH with respect to the minimal connection is
 \bea
 {\cal D}_M^{\mathcal{F}}k^{rs}_P &\defeq& \nabla_M k^{rs}_P +\frac{i}{4}\,\imp\{G^{N_1N_2}{}_M\,\tilde{\pi}^{rs}_{PN_1N_2}\} = -\frac{i}{6}\,F_{MP}{}^{N_1N_2N_3}\,\pi^{rs}_{N_1N_2N_3}
 \cr
 &&-\frac{1}{48}\,\rep \{G^{N_1N_2N_3}\,\tilde{\tau}^{rs}_{MPN_1N_2N_3}\} -\frac{3}{8}\,\rep\{ G_{MP}{}^N\,\tilde{k}^{rs}_N\}
 \cr
 &&+\frac{i}{48}\,g_{MP}\,\imp\{G^{N_1N_2N_3}\,\tilde{\pi}^{rs}_{N_1N_2N_3}\}  +\frac{i}{8}\,\imp\{G^{N_1N_2}{}_{[M}\,\tilde{\pi}^{rs}_{P]N_1N_2}\} ~,
 \label{iibtcfh1}
 \eea
 \bea
 {\cal D}_M^{\cal F} \pi^{rs}_{P_1P_2P_3} &\defeq& \nabla_M \pi^{rs}_{P_1P_2P_3} +\frac{i}{4}\,\imp\{ G_M{}^{N_1N_2}\,\tilde{\tau}^{rs}_{P_1P_2P_3N_1N_2}\} -\frac{3i}{2}\,\imp\{G_{M[P_1P_2}\tilde{k}^{rs}_{P_3]} \}
 \cr
&&+\frac{3}{2}\,\rep\{G^N{}_{M[P_1}\,\tilde{\pi}^{rs}_{P_2P_3]N}\}
\cr
 &&= \frac{i}{8}\,g_{M[P_1}\,F^{N_1\dots N_4}{}_{P_2}\,\tau^{rs}_{P_3]N_1\dots N_4} +\frac{i}{2}\,F^{N_1N_2N_3}{}_{[P_1P_2}\,\tau^{rs}_{P_3M]N_1N_2N_3} \cr
 && -iF_{P_1P_2P_3M}{}^N \,k^{rs}_N + \frac{i}{16}\,\imp\{G^{N_1N_2N_3}\,g_{M[P_1}\tilde{\tau}^{rs}_{P_2P_3]N_1N_2N_3} \} \cr
 && + \frac{i}{4}\,\imp\{G^{N_1N_2}{}_{[M}\tilde{\tau}^{rs}_{P_1P_2P_3]N_1N_2}\}-\frac{3i}{8}\,~g_{M[P_1}\,\imp \{ G_{P_2P_3]}{}^N \tilde{k}^{rs}_N \} \cr
 &&+\frac{i}{2}\, \imp\{ G_{[P_1P_2P_3}\tilde{k}^{rs}_{M]}\} -\frac{1}{48}\,\rep\{{}^{\star}G_{MP_1P_2P_3}{}^{N_1N_2N_3}\,\tilde{\pi}^{rs}_{N_1N_2N_3}\}  \cr
 &&-\frac{3}{8}\,g_{M[P_1}\,\rep\{G_{P_2}{}^{N_1N_2}\,\tilde{\pi}^{rs}_{P_3]N_1N_2}\} -\frac{3}{4}\,\rep\{G^N{}_{[P_1P_2}\,\tilde{\pi}^{rs}_{P_3M]N}\}~,
\label{iibtcfh2}
 \eea
 \bea
 {\cal D}_M^{\cal F}\tau^{rs}_{P_1\dots P_5} &\defeq& \nabla_M \tau^{rs}_{P_1\dots P_5} - 20i\,F^N{}_{M[P_1P_2P_3}\,\pi^{rs}_{P_4P_5]N} + \frac{5}{2}\,\rep\{ G^N{}_{M[P_1}\,\tilde{\tau}^{rs}_{P_2\dots P_5]N}\}
 \cr
 && -\frac{5i}{4}\,\imp\{{}^{\star}G^{N_1N_2}{}_{M[P_1\dots P_4}\,\tilde{\pi}^{rs}_{P_5]N_1N_2}\} - 5i\,\imp\{G_{M[P_1P_2}\,\tilde{\pi}^{rs}_{P_3P_4P_5]}\}
 \cr
 &&= -15i\,F^N{}_{[MP_1P_2P_3}\,\pi^{rs}_{P_4P_5]N} + 10i\, g_{M[P_1}\,F_{P_2P_3P_4}{}^{N_1N_2}\,\pi^{rs}_{P_5]N_1N_2}
 \cr
 &&-\frac{1}{8}\,\rep {}^{\star}G_{P_1\dots P_5M}{}^N\tilde{k}^{rs}_N - \frac{5}{4}\,g_{M[P_1}\rep \{ G_{P_2}{}^{N_1N_2}\,\tilde{\tau}^{rs}_{P_3P_4P_5]N_1N_2}\}
 \cr
 &&-\frac{15}{8}\,\rep\{ G^N{}_{[P_1P_2}\,\tilde{\tau}^{rs}_{P_3P_4P_5M]N}\} + \frac{5}{2}\, g_{M[P_1}\rep\{G_{P_2P_3P_4}\,\tilde{k}^{rs}_{P_5]}\}~,
 \cr
 &&-\frac{15i}{4}\,g_{M[P_1}\,\imp\{G_{P_2P_3}{}^N\,\tilde{\pi}^{rs}_{P_4P_5]N}\} +\frac{5i}{2}\,\imp\{G_{[P_1P_2P_3}\,\tilde{\pi}^{rs}_{P_4P_5M]}\}
 \cr
 && -\frac{5i}{16}\,g_{M[P_1}\,\{{}^{\star}G_{P_2\dots P_5]}{}^{N_1N_2N_3}\,\tilde{\pi}^{rs}_{N_1N_2N_3}\} \cr
 &&+ \frac{9i}{8}\,\imp\{{}^{\star}\,G^{N_1N_2}{}_{[P_1\dots P_5}\,\tilde{\pi}^{rs}_{M]N_1N_2}\} ~,
 \label{iibtcfh3}
 \eea
\bea
 {\cal D}_M^{\cal F}\tilde{k}^{rs}_P &\defeq& \nabla_M\tilde{k}^{rs}_P + i\,Q_M\,\tilde{k}^{rs}_P - \frac{i}{24}\,F_M{}^{N_1\dots N_4}\,\tilde{\tau}^{rs}_{PN_1\dots N_4} + \frac{1}{4}\,\GG_M{}^{N_1N_2}\, \pi^{(rs)}_{PN_1N_2}
 \cr
 &&= -\frac{1}{48}\, \GG^{N_1N_2N_3}\, \tau^{(rs)}_{MPN_1N_2N_3} + \frac{1}{48}\,g_{MP} \,\GG^{N_1N_2N_3}\, \pi^{(rs)}_{N_1N_2N_3}
 \cr
 &&+\frac{1}{8}\,\GG^{N_1N_2}{}_{[M}\, \pi^{(rs)}_{P]N_1N_2} - \frac{3}{8}\,\GG_{MP}{}^N\, k^{(rs)}_N ~,
 \label{iibtcfh4}
 \eea
 \bea
 {\cal D}^{\cal F}_M\tilde{\pi}^{rs}_{P_1P_2P_3} &\defeq& \nabla_M\,\tilde{\pi}^{rs}_{P_1P_2P_3} +i\,Q_M\,\tilde{\pi}^{rs}_{P_1P_2P_3} + \frac{1}{4}\,\GG_M{}^{N_1N_2}\,\tau^{[rs]}_{P_1P_2P_3N_1N_2} \cr
 &&+ \frac{3}{2}\,\GG^N{}_{M[P_1}\,\pi^{[rs]}_{P_2P_3]N} -\frac{3}{2}\,\GG_{M[P_1P_2}\,k^{[rs]}_{P_3]}
 \cr
 &&=\frac{i}{2}\,g_{M[P_1}\,F_{P_2P_3]}{}^{N_1N_2N_3}\,\tilde{\pi}^{rs}_{N_1N_2N_3} - 2i\,F^{N_1N_2}{}_{[P_1P_2P_3}\,\tilde{\pi}^{rs}_{M]N_1N_2}
 \cr
 && - 3i\,F^{N_1N_2}{}_{M[P_1P_2}\,\tilde{\pi}^{rs}_{P_3]N_1N_2} -\frac{1}{48}\,{}^{\star}\GG_{MP_1P_2P_3}{}^{N_1N_2N_3}\,\pi^{[rs]}_{N_1N_2N_3}
 \cr
 && +\frac{1}{16}\,\GG^{N_1N_2N_3}\,g_{M[P_1}\,\tau^{[rs]}_{P_2P_3]N_1N_2N_3} -\frac{1}{4}\,\GG^{N_1N_2}{}_{[P_1}\,^{[rs]}\tau_{P_2P_3M]N_1N_2}
 \cr
 && -\frac{3}{8}\,g_{M[P_1}\,\GG_{P_2}{}^{N_1N_2}\,\pi^{[rs]}_{P_3]N_1N_2} -\frac{3}{4}\,\GG^N{}_{[P_1P_2}\,\pi^{[rs]}_{P_3M]N}
 \cr
 && -\frac{3}{8}\,g_{M[P_1}\,\GG_{P_2P_3]}{}^N\,k^{[rs]}_N + \frac{1}{2}\,\GG_{[P_1P_2P_3}\,k^{[rs]}_{M]}~,
 \label{iibtcfh5}
 \eea
 \bea
 {\cal D}_M^{\cal F}\tilde{\tau}^{rs}_{P_1\dots P_5} &\defeq& \nabla_M\tilde{\tau}^{rs}_{P_1\dots P_5} + i\,Q_M\, \tilde{\tau}^{rs}_{P_1\dots P_5} -10i\,F_{M[P_1\dots P_4}\,\tilde{k}^{rs}_{P_5]} + 5i\, F^{N_1N_2}{}_{M[P_1P_2}\,\tilde{\tau}^{rs}_{P_3P_4P_5]N_1N_2}
 \cr
&&- \frac{5}{4}\, {}^{\star}\GG^{N_1N_2}{}_{M[P_1\dots P_4}\, \pi^{(rs)}_{P_5]N_1N_2} +\frac{5}{2}\,\GG^N{}_{M[P_1}\, \tau^{(rs)}_{P_2\dots P_5]N} - 5\, \GG_{M[P_1P_2}\,\pi^{(rs)}_{P_3P_4P_5]}
\cr
&&=- 5i\, g_{M[P_1}\,F_{P_2\dots P_5]}{}^N\,\tilde{k}^{rs}_N +6i\,F_{[P_1\dots P_5}\,\tilde{k}^{rs}_{M]} -\frac{1}{8}\,{}^{\star}\GG_{P_1\dots P_5M}{}^N\, k^{(rs)}_N
\cr
&&-\frac{5}{4}\,g_{M[P_1}\,\GG_{P_2}{}^{N_1N_2}\, \tau^{(rs)}_{P_3P_4P_5]N_1N_2} -\frac{15}{8}\,\GG^N{}_{[P_1P_2}\, \tau^{(rs)}_{P_3P_4P_5M]N}
\cr
&&-\frac{15}{4}\,g_{M[P_1}\,\GG_{P_2P_3}{}^N\, \pi^{(rs)}_{P_4P_5]N} + \frac{5}{2}\,\GG_{[P_1P_2 P_3}\, \pi^{(rs)}_{P_4P_5M]} +\frac{5}{2}\,g_{M[P_1}\,\GG_{P_2 P_3 P_4}\, k^{(rs)}_{P_5]}
\cr
&& -\frac{5}{16}\,g_{M[P_1}{}^{\star} \GG_{P_2\dots P_5]}{}{}{}^{N_1N_2N_3}\, \pi^{(rs)}_{N_1N_2N_3} + \frac{9}{8}\,{}^{\star}\GG^{N_1N_2}{}_{[P_1\dots P_5}\, \pi^{(rs)}_{M]N_1N_2}~,
\label{iibtcfh6}
 \eea
where we have not made a sharp distinction between spacetime and frame indices.

Following the same prescription as in the AdS backgrounds and after decomposing  the form bilinears  into the real and the imaginary parts, one finds that the (reduced) holonomy of the TCFH connection is included in (the connected component of) $SO(9,1)\times GL(518)\times GL(496)$. This result agrees with the calculation in \cite{lgjpgp} performed in the string frame.

\end{appendices}

\end{document}